\documentclass[10pt]{revtex4-1}
\usepackage{color}

\usepackage{graphicx}
\usepackage{amssymb,amsthm}
\usepackage{enumerate}

\usepackage{amsmath}

\usepackage{graphicx}%
\usepackage{dcolumn}%
\usepackage{bm}%

\usepackage{caption}
\usepackage{subcaption}
\usepackage{float}

\usepackage[breaklinks]{hyperref}
\usepackage{url}
\usepackage{breakurl}

\newcommand{\beq}[0]{\begin{equation}}
\newcommand{\eeq}[0]{\end{equation}}

\newcommand{\thet}{\vartheta}

\newcommand{\ds}{\displaystyle}

\newcommand{\pa}{\partial}

\newcommand{\ud}{{\mathrm d}}
\newcommand{\bc}{\begin{center}}
\newcommand{\ec}{\end{center}}
\newcommand{\lesq}{\leqslant}

\begin{document}
\title{Chaotic behavior of  three interacting vortices in a confined Bose-Einstein condensate}
\author{Nikos \surname{Kyriakopoulos}}
\affiliation{SUPA, Department of Physics and Institute for Complex Systems and Mathematical Biology, King's College, University of Aberdeen, Aberdeen, AB24 3UE, UK}
\author{Vassilis \surname{Koukouloyannis}}
\affiliation{Physics Department, Aristotle University of Thessaloniki, GR-54124, Thessaloniki, Greece}
\author{Charalampos \surname{Skokos}}
\affiliation{Department of Mathematics and Applied Mathematics, University of Cape Town, Rondebosch, 7701, South Africa}
\affiliation{Physics Department, Aristotle University of Thessaloniki, GR-54124, Thessaloniki, Greece}
\author{Panayotis \surname{Kevrekidis}}
\affiliation{Department of Mathematics and Statistics, University of Massachusetts, Amherst MA 01003-9305, USA}

\begin{abstract}
Motivated by recent experimental works, we investigate a system of vortex dynamics in an atomic Bose-Einstein condensate (BEC), consisting of three vortices, two of which have the same charge. These vortices
are modeled as a system of point particles which possesses a Hamiltonian structure. This tripole system constitutes a prototypical model of vortices in BECs exhibiting chaos. By using the angular momentum integral of motion we reduce the study of the system to the investigation of a two degree of freedom Hamiltonian model and acquire quantitative results about its chaotic behavior. Our investigation tool is the construction of scan maps by using the Smaller ALignment Index (SALI) as a chaos indicator. Applying this approach to a large number of initial conditions we manage to accurately and efficiently measure the extent of chaos in the model and its dependence on physically important parameters like the energy and the angular momentum of the system.
\end{abstract}

\maketitle

\section {Introduction}

The study of two-dimensional particle dynamics resulting from a
logarithmic interaction potential is a theme of broad and diverse
interest in Physics. Arguably, the most canonical example of both
theoretical investigation and experimental relevance is the
exploration of fluid and superfluid
vortex patterns and crystals, as is evidenced
e.g. by the review of Aref et al.~\cite{aref1} and the book of Newton~\cite{newton1}.
However, numerous additional examples ranging from
electron columns in
Malmberg-Penning
traps~\cite{fajans} to magnetized, millimeter sized
disks rotating at a liquid-air interface~\cite{whitesides,whitesides2}
are also characterized by the same underlying mathematical
structure and hence present similar dynamical features.

In recent years, the field of atomic Bose-Einstein condensates
(BECs)~\cite{book1,book2}
has offered an ideal playground for the realization of a diverse
host of configurations showcasing remarkable vortex patterns and
dynamics. The early efforts along this direction principally
focused on the existence and dynamical robustness/stability properties
of individual vortices (including multi-charge ones that were
generically identified as unstable in experiments), as well as of
large scale vortex lattices created upon suitably fast
rotation~\cite{fetter1,fetter2,chamoun,book3}.
Some of the early theoretical and experimental
efforts also touched upon few-vortex
crystals~\cite{castin,chevy}. Yet, it was not until the development
of more recent experimental techniques, such as the minimally
destructive imaging~\cite{freilich,dsh2,dsh3},
the imaging of dragged laser beams through the BEC~\cite{bpa10},
or the quadrupolar excitations spontaneously producing multi-vortex
states~\cite{bagn}
that few-vortex dynamics drew
a sharp focus of the research effort. It is worthwhile to note that
in this BEC context, some of the standard properties and conservation
laws of the vortex system~\cite{arefjmp} still apply, including e.g.
the angular momentum (i.e., the sum of the squared distances of the vortices
from the trap center multiplied by their respective topological charge)
or the Hamiltonian of the vortex system. However, others such as the
linear momentum are no longer preserved. This is due to the local
vortex
precession term arising in the dynamics as a result of the presence of the
external (typically parabolic) trap~\cite{fetter1,fetter2}.

Motivated by the ongoing experimental developments, and perhaps
especially the work of Seman et al.~\cite{bagn},
in the recent work of Koukouloyannis et al.~\cite{tripole2013}, a detailed study of the transition from regular to progressively chaotic behavior has been performed in the tripole configuration (consisting of two vortices of one
circulation and one of the opposite circulation). This has been
achieved by using a
sequence of Poincar\'e sections with the angular momentum $L$ of the vortex system as a parameter. Notice that while this tripole system without the
local BEC-trap induced precession is
integrable
(see e.g.~the discussion of Aref and co-workers~\cite{arefjmp,arefone}),
here the absence of
linear momentum conservation renders chaotic dynamics accessible
at this level.
In this context the main bifurcations which lead to the destabilization of the system and the eventual appearance of chaotic behavior have been observed. Our
aim in the present work is to provide more quantitative results about the
chaotic behavior of the system for various energy levels. As a principal
tool to this effect, we will employ an efficient  chaos detection method, the so--called Smaller ALignment Index (SALI).

Our study is structured as follows. In section II, we briefly present the
setup of the theoretical particle model developed earlier~\cite{dsh2,dsh3,tripole2013},  which we will
 use in the present study. In section III, we present the numerical tools that we use in this work, namely the chaoticity index SALI and the scan maps that can be derived by using this index. 
After that, in section IV.A we perform an extended dynamical study of the system for a typical value of its energy $h$ by using its angular momentum $L$ as a parameter. In this study we concentrate mainly in the study of the evolution of the permitted area of motion and the chaoticity of the system as the value of $L$ varies. In addition, based on some physical properties of our system we argue that SALI is a more relevant tool of investigation for this study
than the maximum Lyapunov exponent (mLE). In section IV.B we generalize this study in order to acquire a more global picture of the dynamics of the system by including most of the physically meaningful values of the energy of the system.  Finally, we summarize our findings and present some directions for future study in section V.

\section{The model}
In this section we briefly present the model used also in Koukouloyannis et al.~\cite{tripole2013}for the study of the dynamical behavior of a system of three interacting vortices in quasi-two-dimensional (pancake shaped) BEC. We consider two of them having the same topological charge  $S_1=S_3=1$ while the third has  $S_2=-1$, following the experimental results of Seman et al.~\cite{bagn}. In this case, if the vortices are well-separated, they can be considered as point quasi-particles and the corresponding normalized equations describing their motion are
\begin{equation}
\begin{array}{rrl}
\dot{x}_i=&\ds-S_i\frac{y_i}{1-r_i^2}-c\sum_{j=1, j\neq i}^NS_j\frac{y_i-y_j}{r_{ij}^2}& \\
\dot{y}_i=&\ds S_i\frac{x_i}{1-r_i^2}+c\sum_{j=1, j\neq i}^NS_j\frac{x_i-x_j}{r_{ij}^2}&\qquad i,j=1, 2,3.
\end{array}
\label{nomal_eqs}
\end{equation}
where $(x_i, y_i)$ stand for the coordinates of the $i$-th vortex in the plane of motion, while $r_i=\sqrt{x_i^2+y_i^2}$ and $r_{ij}=\sqrt{\left(
x_{i}-x_{j}\right) ^{2}+\left( y_{i}-y_{j}\right) ^{2}}$. The parameter $c$ is connected to the physical properties of the BEC and a typical value for it has been estimated e.g.~by Navarro et al.~\cite{dsh3}
to be $c \approx 0.1$. The above equations have been rescaled so the Thomas-Fermi radius of the BEC (which characterizes the radial extent of the BEC) is $R_{TF}=1$.  Consequently, $0\lesq r_i<1$.

This system can be described by a three degrees of freedom Hamiltonian, where each pair  of coordinates $(x_i, y_i)$ corresponds to one degree of freedom. The above equations of motion, for the particular choice of $S_1$, $S_2$, $S_3$  can be derived by the Hamiltonian
\beq H=\frac{1}{2}\sum_{i=1}^3\ln(1-r_i^2)+\frac{c}{2}\left[\ln(r_{12}^2)-\ln(r_{13}^2)+\ln(r_{23}^2)\right]\label{ham}\eeq
via the canonical equations $\dot{x}_i=S_i\frac{\pa H}{\pa y_i}$, $\dot{y}_i=-S_i\frac{\pa H}{\pa x_i}$.
Considering ${\bf q}=(x_1,\ y_2,\ x_3)$, ${\bf p}=(y_1,\ x_2,\ y_3)$ we acquire the usual form of the canonical equations $\dot{q}_i=\frac{\pa H}{\pa p_i}$, $\dot{p}_i=-\frac{\pa H}{\pa q_i}$ for the system's evolution.

Applying two successive canonical transformations: a) $(x_i,y_i)\mapsto (w_i,R_i)$ defined by
\beq q_i=\sqrt{2R_i}\sin(w_i)\ ,\quad p_i=\sqrt{2R_i}\cos (w_i),\quad i=1\ldots3\label{can_transf_1},\eeq
and b) $(w_i,R_i)\mapsto (\phi_{1,2},\thet,J_{1,2},L)$ according to
\beq\begin{array}[t]{rlcrl}
\phi_1=&w_1-w_3&\quad\quad&J_1=&R_1\\
\phi_2=&w_2+w_3&\quad\quad&J_2=&R_2\\
\thet=&w_3&\quad\quad&L=&R_1-R_2+R_3,
\end{array}\label{can_transf_2}\eeq
the Hamiltonian \eqref{ham} assumes the form
\beq\begin{array}[t]{rcl}
H&=&\ds\frac{1}{2}\left[ \ln(1-2J_1)+\ln(1-2J_2)+\ln(1-2(L-J_1+J_2)) \right]\\[10pt]
 &+&\ds\frac{c}{2}\left[\ \ln(4J_2-2J_1+2L-4\sqrt{J_2}\sqrt{L-J_1+J_2}\sin(\phi_2))\right.\\[10pt]
& & \ds\quad\ -\ln(2L+2J_2-4\sqrt{J_1}\sqrt{L-J_1+J_2}\cos(\phi_1)) \\[10pt]
& & \ds\quad\ \left. +\ln(2J_1+2J_2-2\sqrt{J_1}\sqrt{J_2}\sin(\phi_1+\phi_2))\ \right].\\[10pt]
\end{array}\label{ham_red}\eeq
Since the above Hamiltonian is autonomous, the {\it energy} of the system, which is expressed by $H$, is conserved. In addition, $\thet$ is ignorable and consequently its conjugate generalized momentum $L$ \eqref{can_transf_2}, the {\it angular momentum} of the system, is also an integral of motion. Thus, Hamiltonian \eqref{ham_red} can be considered as a two degrees of freedom system with $L$ as a parameter.

In what follows we use the value of the energy $h$ of the system and the value of the angular momentum $L$ as the main parameters of our study. Both values  depend on the particular vortex  configuration, i.e.~the set of initial conditions of each orbit as $h=H(x_{10},y_{10},x_{20},y_{20},x_{30},y_{30})$ and $L=L(x_{10},y_{10},x_{20},y_{20},x_{30},y_{30})$.

Before we present our main results, we will briefly discuss the numerical methods of this study.


\section{Numerical Methods}

\subsection{The Smaller ALignment Index - SALI}

The most commonly used chaos indicator is the computation of the maximum Lyapunov exponent (mLE)~\cite{BGGS_80a,BGGS_80b,S10}, which is based on the evolution of one deviation vector from the studied orbit. The main drawback for using the mLE is the long time needed for the index to converge to its limiting value, especially for chaotic orbits that stick close to regular ones for long times.

Many methods have been developed over the years that overcome this problem and allow the fast and reliable characterization of orbits as chaotic or regular, like the Fast Lyapunov Indicator (FLI)~\cite{FLI,FLI2} and its variants~\cite{Barrio,Barrio2}, the Smaller (SALI)~\cite{sali} and the Generalized (GALI)~\cite{gali} ALignment Indices, the Mean Exponential Growth of Nearby Orbits (MEGNO)~\cite{megno,megno2}, the Relative Lyapunov Indicator (RLI)~\cite{RLI,RLI2}, the Frequency Map Analysis~\cite{Laskar,Laskar2}, the `0-1¢ test~\cite{01,012}, and the Covariant Lyapunov Vectors (CLV) method~\cite{CLV,Wolfe}. A concise presentation of some of these methods, as well as a comparison of their performances can be found in the works of Maffione et al.\cite{MDCG11} and Darriba et al.\cite{DMCG12}. In our study we will use the SALI method, which proved to be an efficient indicator of chaos.
 The SALI depends on the evolution of two initially different deviation vectors, which are repeatedly normalized from time to time  and checks whether they will align (chaotic orbit) or not (regular orbit). It has been shown that SALI tends exponentially fast to zero for chaotic orbits, while it fluctuates around constant, positive values for regular ones\cite{SABV03,SABV04}. In practice, we require SALI to become smaller than a very small threshold value (in our study we set $\text{SALI}_{thres}=10^{-12}$) to characterize an orbit as chaotic. The different behavior of the SALI for chaotic and regular orbits makes it an efficient chaos indicator, as its many applications to a variety of dynamical systems\cite{SESS04,BS06,ABS06,CLMV07,MSCHJD07,SHC09,ABB10,MA11,BCSV12,BCSPV12,BLBB13,ABDNT13} illustrate. Thus, SALI constitutes an ideal numerical tool for the purposes of our study, as its computation for a large sample of initial conditions allows the construction of phase space charts (which we will call `scan maps') where regions of chaoticity and regularity are clearly depicted and identified.

\subsection{The scan map}

\begin{center}
\begin{figure}[h]
	\begin{subfigure}{0.28\textwidth}
	\caption{$L=-0.25$}
	\includegraphics[width=\textwidth]{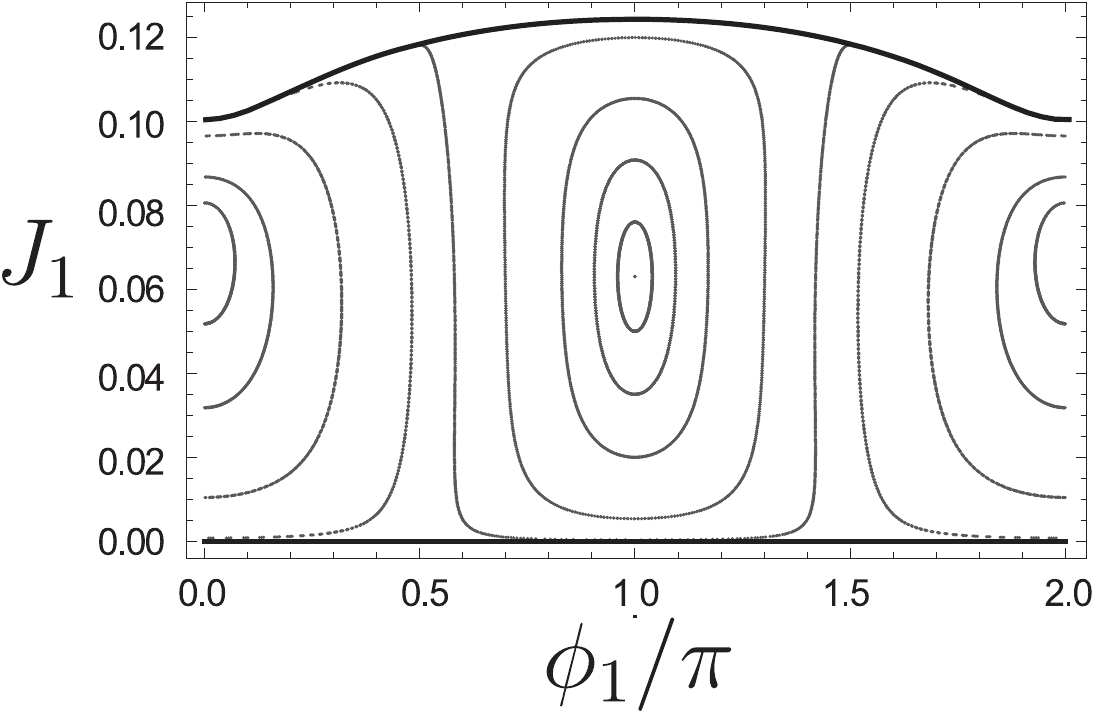}
	\includegraphics[width=\textwidth]{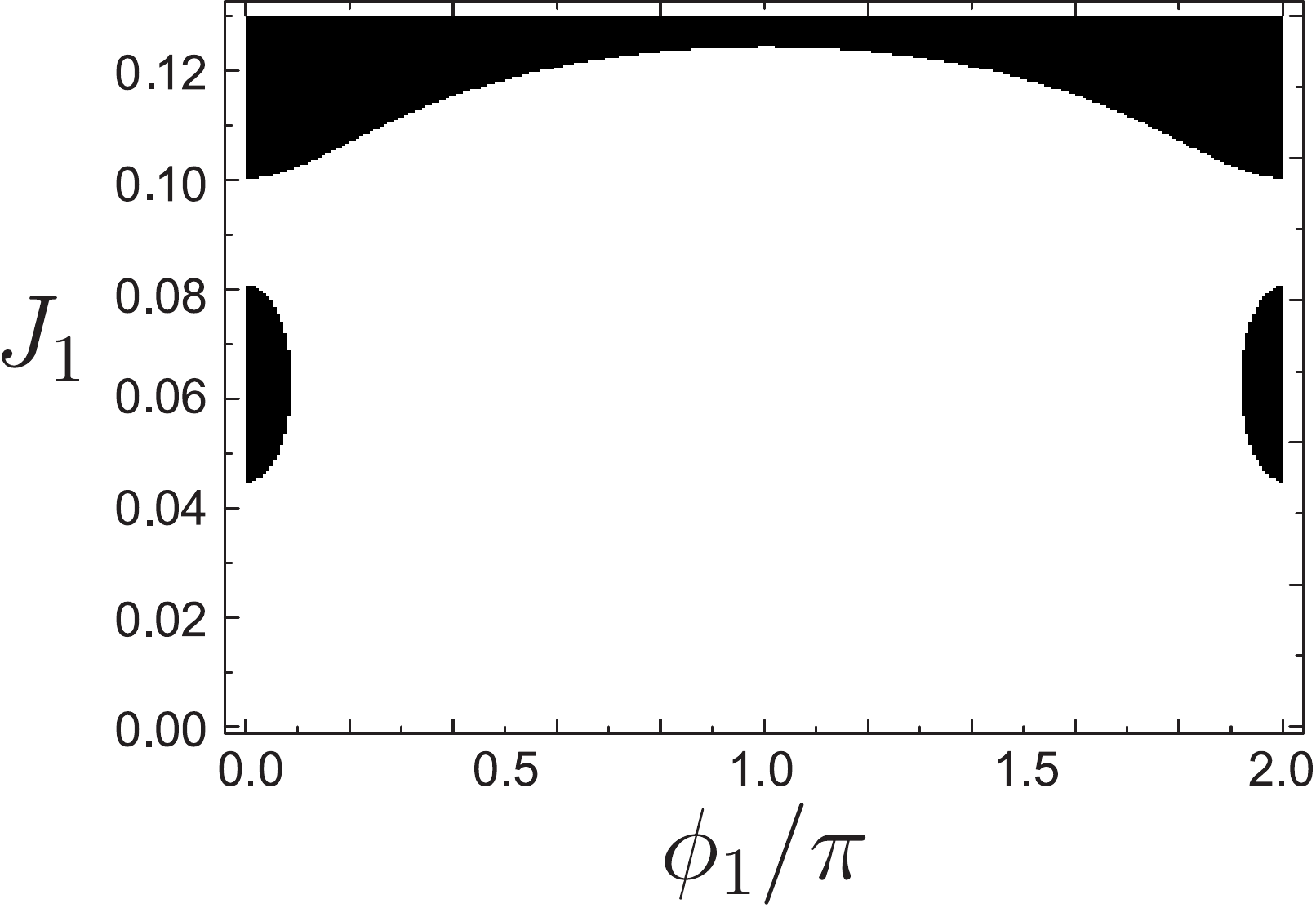}
	\end{subfigure}
	\begin{subfigure}{0.28\textwidth}
	\caption{$L=-0.05$}
	\includegraphics[width=\textwidth]{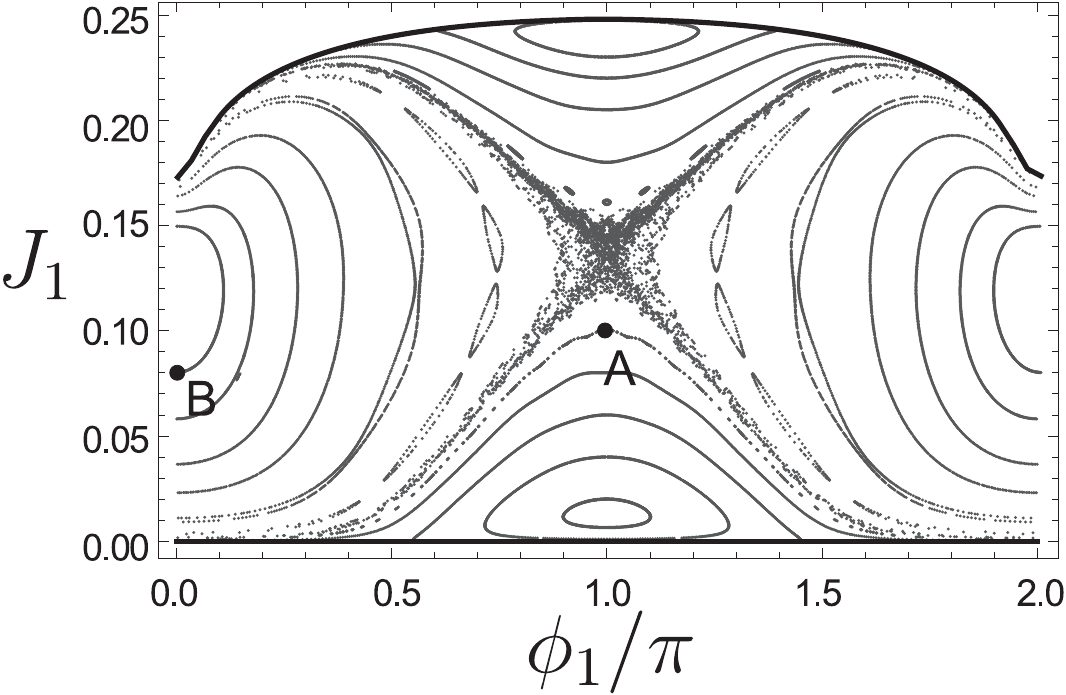}
	\includegraphics[width=\textwidth]{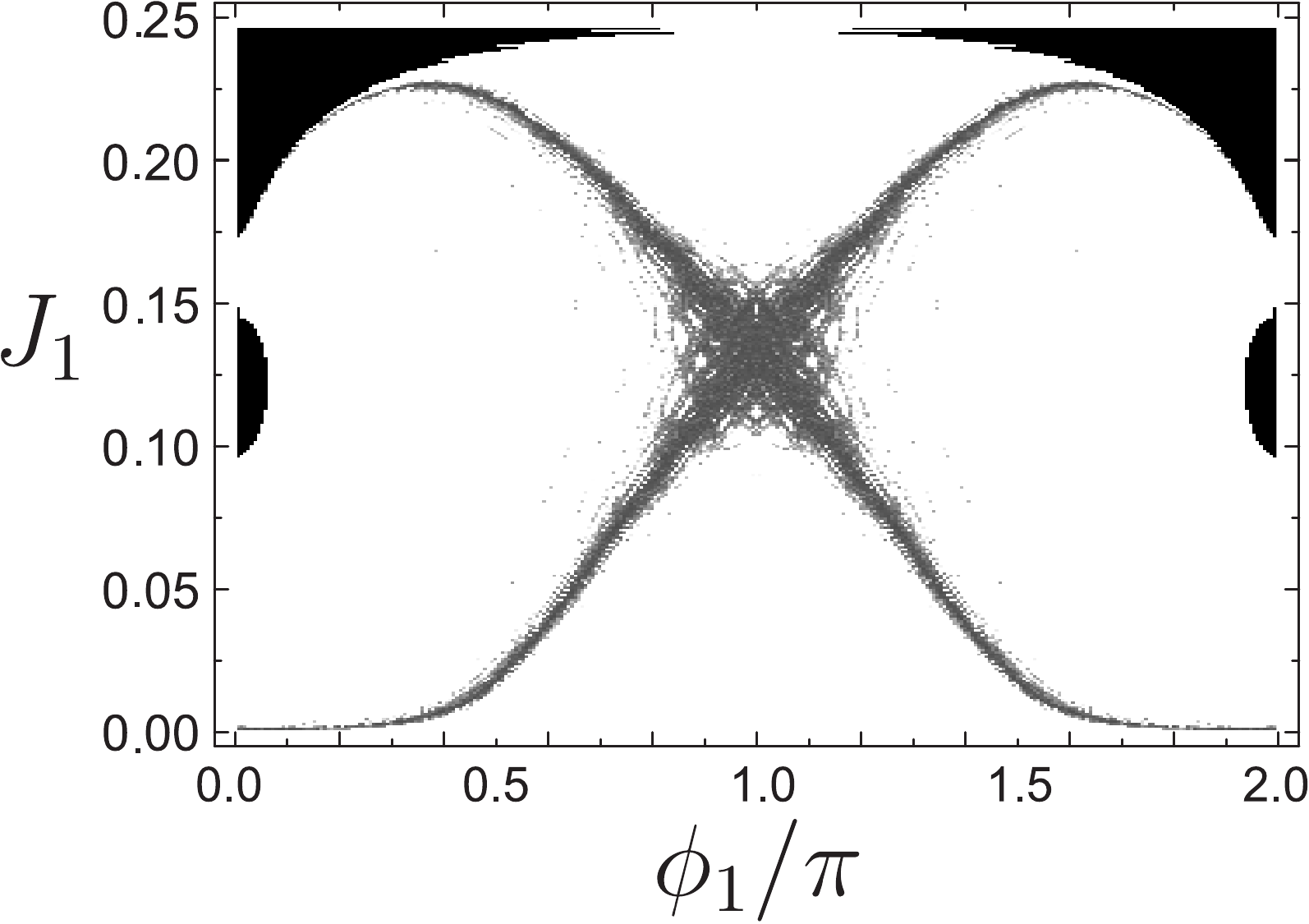}
	\end{subfigure}
	\begin{subfigure}{0.28\textwidth}
	\caption{$L=0.05$}
	\includegraphics[width=\textwidth]{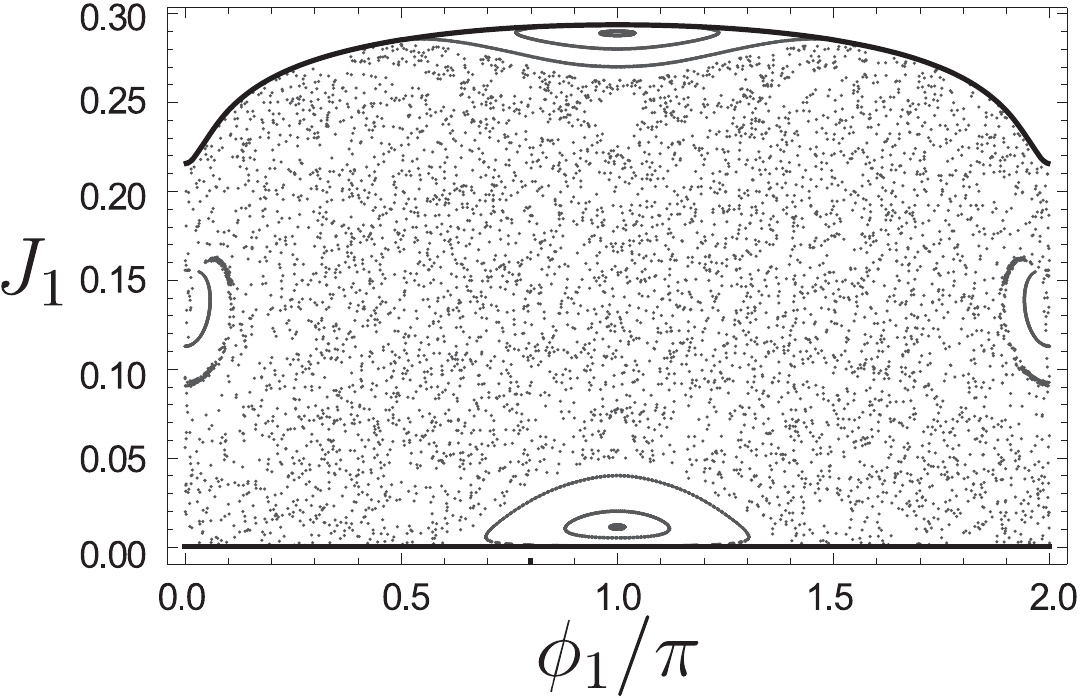}
	\includegraphics[width=\textwidth]{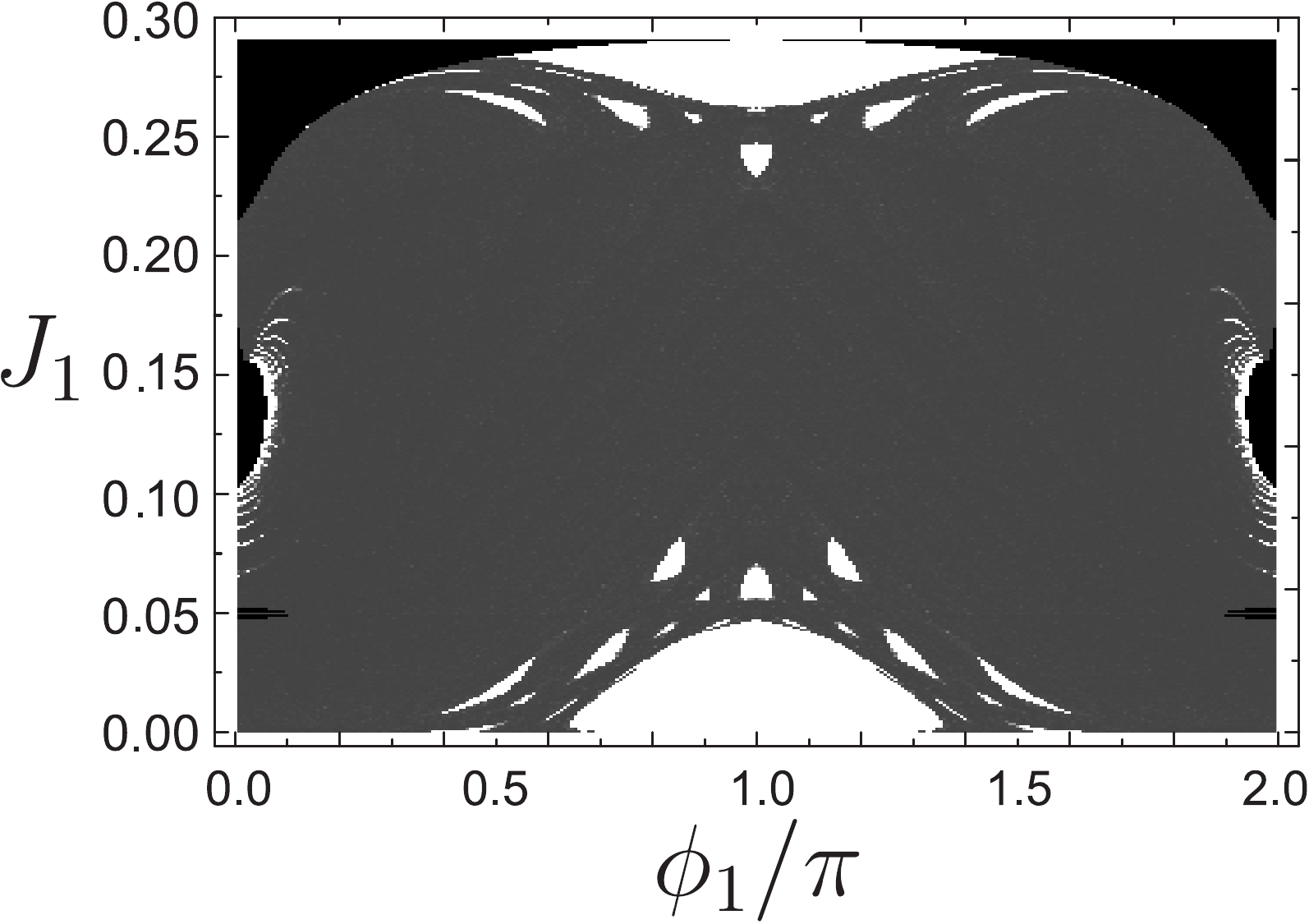}
	\end{subfigure}

	\begin{subfigure}{0.28\textwidth}
	\caption{$L=0.25$}
	\includegraphics[width=\textwidth]{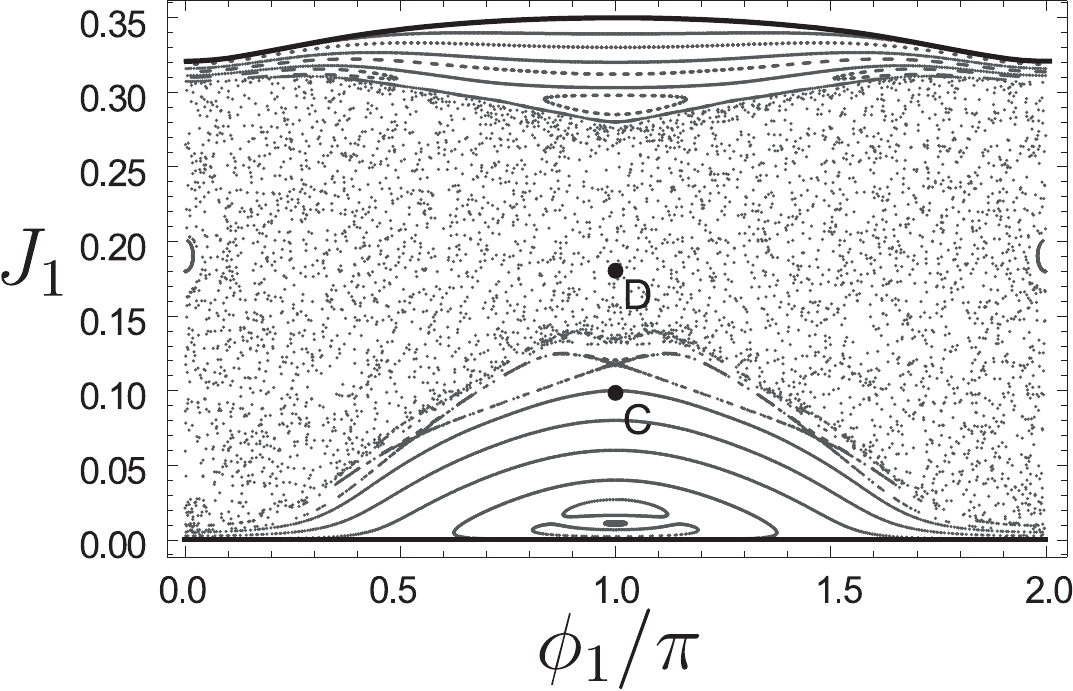}
	\includegraphics[width=\textwidth]{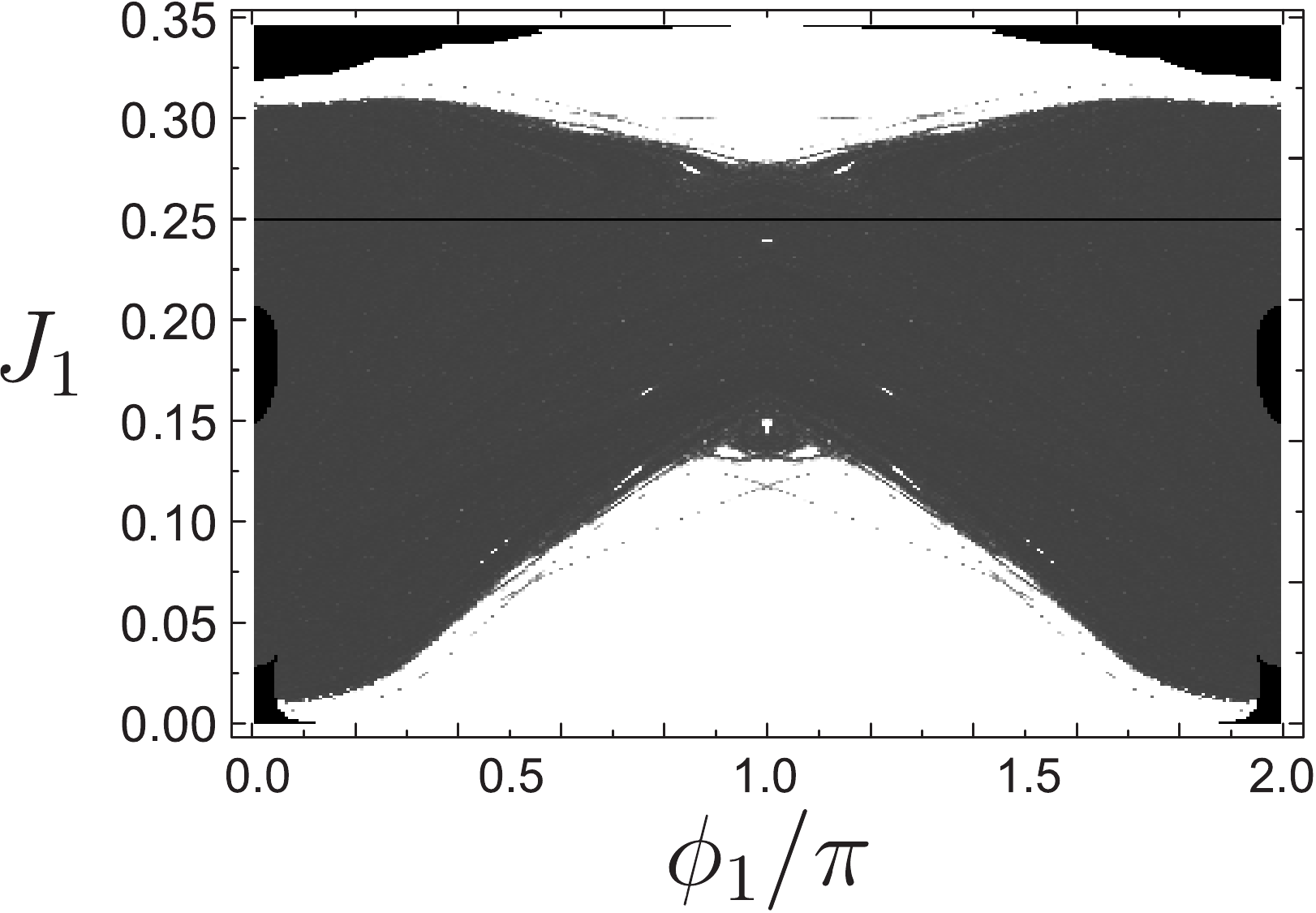}
	\end{subfigure}	
	\begin{subfigure}{0.28\textwidth}
	\caption{$L=0.45$}	
	\includegraphics[width=\textwidth]{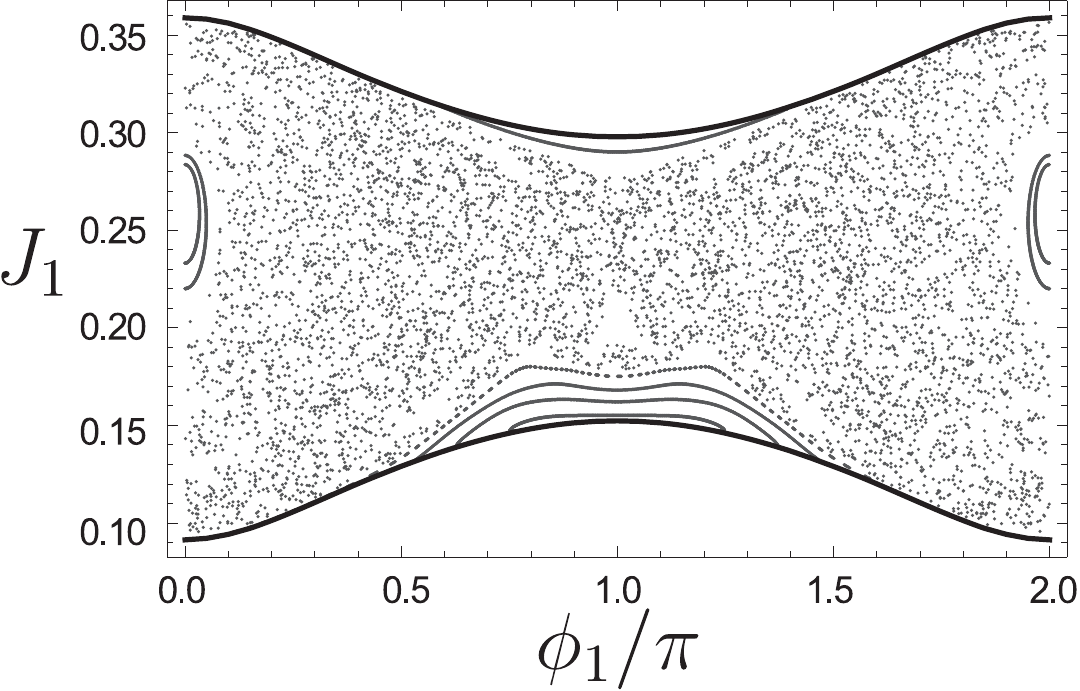}
	\includegraphics[width=\textwidth]{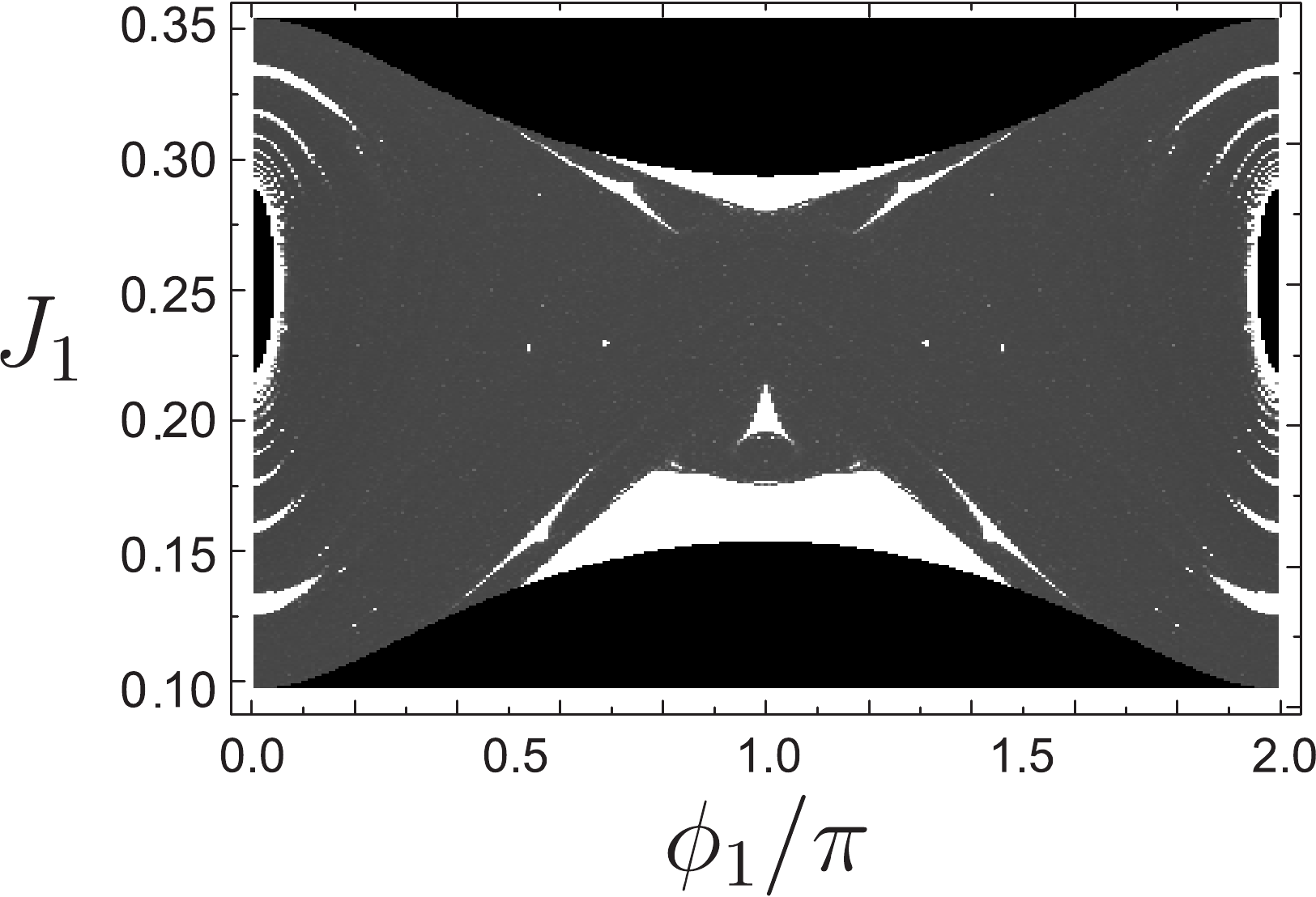}
	\end{subfigure}
	\begin{subfigure}{0.28\textwidth}
	\caption{$L=0.50$}
	\includegraphics[width=\textwidth]{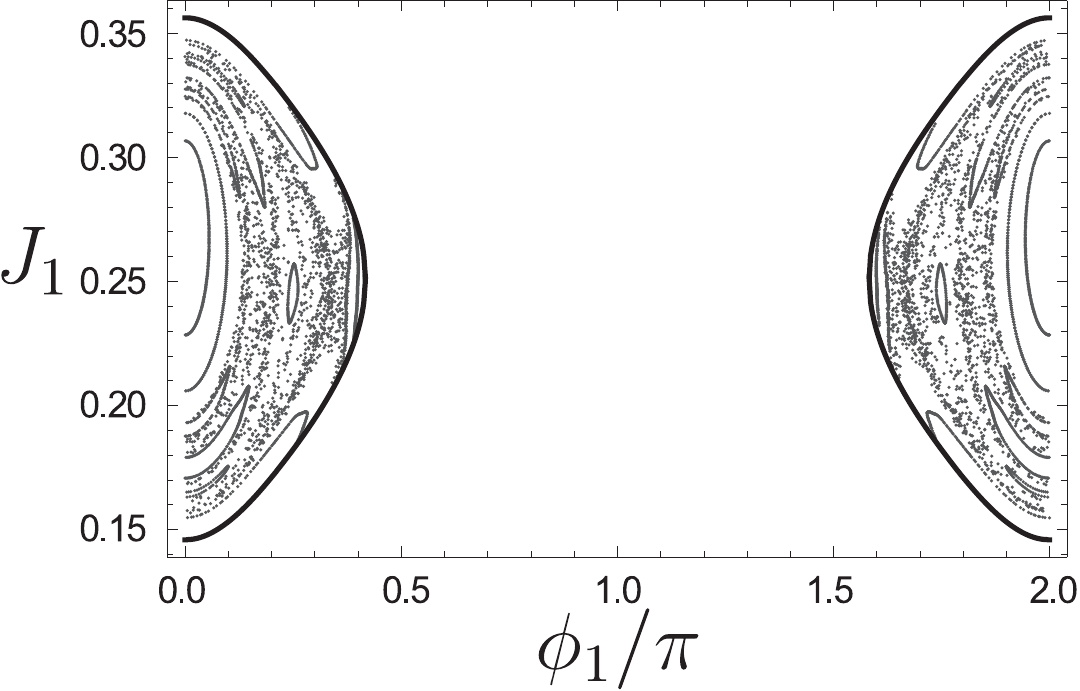}
	\includegraphics[width=\textwidth]{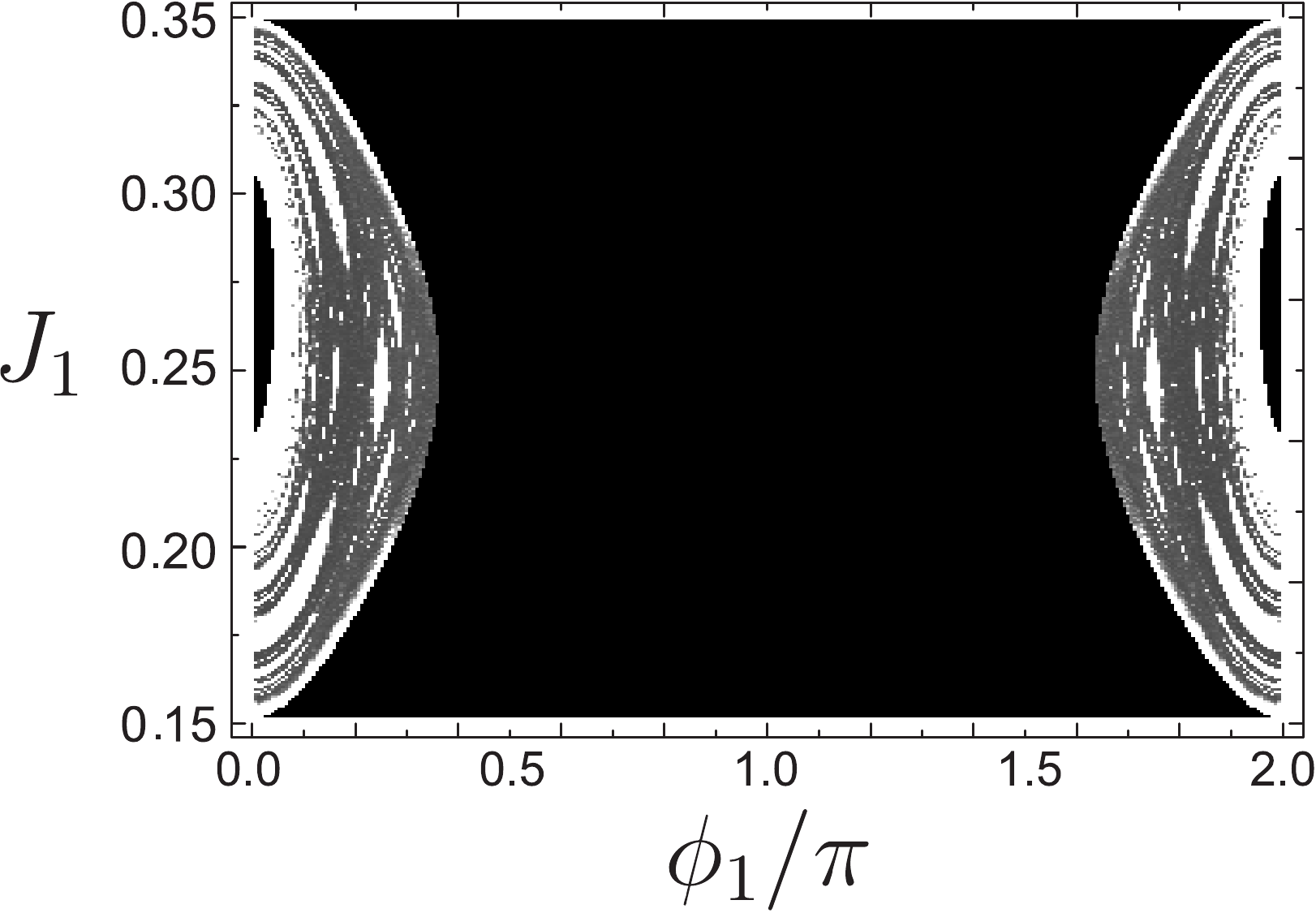}
	\end{subfigure}

	\begin{subfigure}{\textwidth}
	\includegraphics{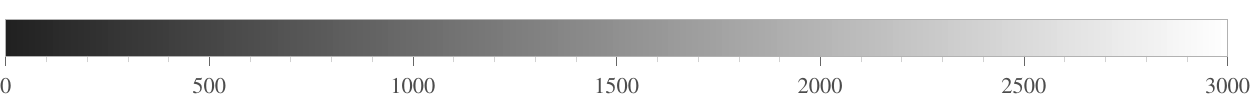}
	\end{subfigure}

\caption{Poincar\'e surfaces of section (upper panels) and the corresponding SALI scan maps (lower panels) of Hamiltonian \eqref{ham_red} for energy $h=-0.7475$ and varying values of $L$. In the PSSs the thick black curves correspond to boundaries of motion. In the scan maps the black areas correspond to non-permitted orbits. The gray scale shown below the panels is used for coloring each permitted initial condition according to its $t_{S_0}$ value (see text for more details). So, dark colored points correspond to orbits with small $t_{S_0}$ (chaotic orbits) and light colored points correspond to orbits with large $t_{S_0}$ (regular orbits). In the upper plots of panels (b) and (d) the initial conditions of some particular orbits studied in Sect.~\ref{sub:h0.7475} are also indicated.}
\label{fig: sections_scans}
\end{figure}
\end{center}

In order to calculate a scan  map we first have to define a Poincar\'e surface of section (PSS)~\cite{lichtenberg}.  Since our Hamiltonian is considered to be a two degrees of freedom one with $L$ as a parameter, for the PSS to be defined we have to consider fixed values $h$ and $L$ for the energy and the angular momentum respectively. We also consider a constant value for $\phi_2$, namely $\phi_2=\pi/2$. In this way the plane $(\phi_1, J_1)$ is defined as the plane of the PSS and $J_2$ is calculated at each point of the section by (\ref{ham_red}).
Note that the value of $\phi_2=\pi/2$ corresponds to the configuration where the $S_2$ and $S_3$ vortices lie on the half-line having the center of
the condensate on its edge as can be seen from the transformations \eqref{can_transf_1} and \eqref{can_transf_2}. The main motion of the vortices is dictated by their gyroscopic precession which has as a result vortices with opposite charge to rotate in different directions. Consequently, as we can see from (\ref{can_transf_2}), the angle $\phi_2$ will take almost all the values, independently of the choice of the specific orbit. Thus, the section which corresponds to $\phi_2=\pi/2$ is appropriate  for revealing the system's main dynamical features as it is crossed by the vast majority of the permitted orbits.
Several PSSs obtained by this approach, for $h=-0.7475$ and various values of $L$ are seen in the upper panels of  \autoref{fig: sections_scans}.

In order to construct a scan map, like the ones shown in the lower panels of \autoref{fig: sections_scans}, we select an equally spaced  grid of $300\times300$ initial conditions $(\phi_1,J_1)$ on the PSS and compute SALI for each  orbit\footnote{We note that the integration of the orbit and of the two deviation vectors needed for the computation of the SALI is done by using the DOPRI853 integrator\cite{hairer}}. When the value of SALI becomes $\text{SALI}<\text{SALI}_{thres}=10^{-12}$ we consider SALI to practically be zero  and the corresponding orbit to be chaotic. We denote the time needed for an orbit to reach this threshold  $t_{S_0}(\phi_{10},J_{10})$. The maximum integration time we consider is $t_{max}=3000$. If $\text{SALI}(t_{max})>\text{SALI}_{thres}$ then the orbit is considered to be regular. In that case, we set  $t_{S_0}$ to be $t_{S_0}=t_{max}$. Depending on the value of $t_{S_0}(\phi_{10},J_{10})$, we assign a color to each point of the grid. In particular, darker colored points correspond to orbits with smaller $t_{S_0}$, while lighter colored points correspond to orbits with larger $t_{S_0}$. In this way we construct color charts of the PSS  based on how fast the chaotic nature of an orbit is revealed. These scan maps clearly show not only the regions where regular and chaotic motion occurs, as the comparison with the PSS plots in the upper panels of \autoref{fig: sections_scans} easily verifies, but also indicate regions with different degrees of chaoticity. Finer grids and longer integration times were also considered, but the results they provided were not significantly different from the ones presented in \autoref{fig: sections_scans}, while the additional computational time required was extremely longer. Hence, the choice of the $300\times300$ grid and the value $t_{max}=3000$ have been deemed to be the most efficient in order to reveal the details of the dynamical behavior of this system.


\section{Results}

\subsection{Dynamical behavior of the system for h=-0.7475}
\label{sub:h0.7475}

The dynamical behavior of Hamiltonian (\ref{ham_red}) has been studied in Koukouloyannis et al.\cite{tripole2013} for the value of the energy $h=-0.7475$ and increasing values $L$ of the angular momentum. This behavior is summarized in \autoref{fig: sections_scans}, where various PSSs are shown together with the corresponding scan maps. This value of $h$ refers  to a `typical' configuration of the system i.e.~a configuration where the vortices are well separated and not close to the Thomas-Fermi radius. As it can be seen from (\ref{can_transf_2}), since $0<R_i<0.5$ the typical range of the values of $L$ is $-0.5<L<1$. But, since the energy constraint must also be fulfilled, the range is actually smaller. In particular, for $h=-0.7475$  the range considered is $-0.45\lesq L\lesq0.55$.

For values $L<-0.218$ the system is fully organized featuring only regular orbits as it can be seen in \autoref{fig: sections_scans}(a). For a critical value of $L\simeq-0.218$ the central periodic orbit  destabilizes through a pitchfork bifurcation, and a chaotic region is subsequently created (\autoref{fig: sections_scans}(b)). This region gets wider as $L$ increases (\autoref{fig: sections_scans}(c)). For even larger values of $L$, the permitted area of the PSS shrinks, as we will see in detail later on, (\autoref{fig: sections_scans}(e)--(f)) and finally all the allowed configurations of the system correspond to regular orbits which are concentrated around an
orbit involving the collision of the vortices $S_1 - S_3$.

In the corresponding scan maps we  observe some black areas which represent rejected initial conditions of the grid. There are three reasons to reject an initial condition on the PSS. The first is that the specific point does not comply with both the energy and angular momentum constraints of the system. These are the upper and lower black areas in the lower panels of \autoref{fig: sections_scans}(a)--(e) and the large, central, black area of \autoref{fig: sections_scans}(f).
The second reason to exclude an initial condition is if a particular configuration corresponds to a collision orbit, i.e. two vortices lie at the same point of the configuration space $(x-y)$. This state is meaningless both physically and mathematically, since the energy of the system becomes infinite. This case is visible in \autoref{fig: sections_scans}(d) where a horizontal black line is shown at $J_1=0.25$, which corresponds to a collision between $S_2$ and $S_3$. The third reason is
purely physical: if the initial condition represents a configuration in which the two co-rotating vortices $S_1$ and $S_3$ lie close to each other, these two vortices become `trapped' in a motion where they rotate around each other. This is called the `satellite' regime. Additionally, if the distance between them is too small ($r_{13}<0.1$), our model does not describe the dynamics accurately, as it was constructed under the assumption that vortices behave like particles retaining their structure unchanged, which of course is not true when they acquire this level of proximity. So in our study we do not try  to tackle questions related to close encounters of the vortices. This restriction corresponds to the small black areas on the left and right end sides of the scan maps. In this consideration we have not excluded the cases where the counter-rotating vortices come close to each other since in this case they are not trapped but instead they just pass by each other and continue their motion.

In this work we are interested, not only to see the general dynamical behavior of the system, but in acquiring more quantitative results, than the ones described above, concerning the permitted area of motion and the chaoticity percentages of the system.

\subsubsection{Permitted area of motion}
\label{area}
The boundaries of the permitted areas in \autoref{fig: sections_scans} are calculated by the requirement that one of the vortices will pass through the origin \cite{tripole2013} ($R_i=0$). From the transformation \eqref{can_transf_2} we see that $R_2$ has a negative contribution to $L$, while $R_1$ and $R_3$ contribute positively. Thus, for low values of $L$, the $S_2$ vortex is moving away from the origin and the boundaries are determined by the $R_1=0$ and $R_3=0$ constraints. In particular, the former condition provides the $J_{1_{lo}(R_1=0)}=0$ boundary, while the latter gives $J_2=J_1-L$. Since in each panel of \autoref{fig: sections_scans} we consider fixed values for $h$ and $L$ and in addition we set $\phi_2=\pi/2$ for the construction of the PSSs, Hamiltonian \eqref{ham_red} provides an implicit relation $J_{1_{up}(R_3=0)}=J_{1_{up}}(\phi_1;h,l)$ for the upper boundary of the permitted area. On the other hand, for high values of $L$ the $J_{1_{lo}(R_2=0)}$ and $J_{1_{up}(R_2=0)}$ boundaries are both calculated by the constraint $R_2=0$, through similar considerations. 

\begin{figure}
	\centering
		\includegraphics[width=8cm]{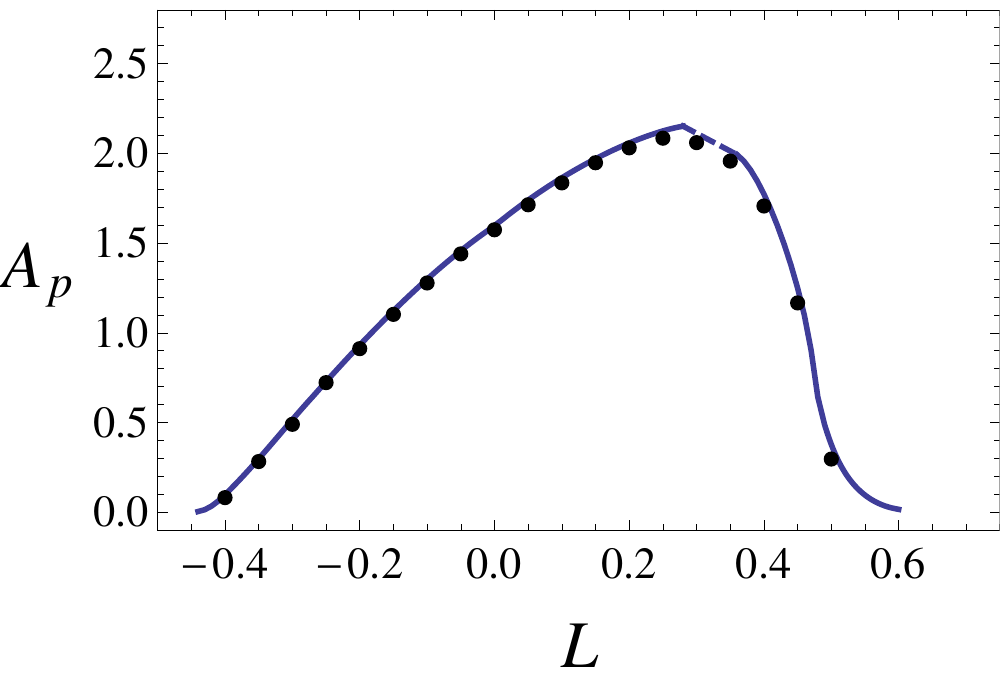}
	\caption{The permitted area of motion for $h=-0.7475$ 
shown as a function of $L$. The solid line represents the calculation of the area using \eqref{eq:area_int}. The dots represent the results obtained by estimating the size of the permitted area as the sum of the areas of all the small rectangles of the grid of the PSS which correspond to permitted orbits. The dashed line connecting the two parts of the solid curve corresponds to the region of the values of $L$ in which there is an ambiguity in the calculation of the area through \eqref{eq:area_int} discussed in the text.}
	\label{fig:area_h_0_7475_nt2}
\end{figure}

The permitted area can now be numerically calculated by the integral
\begin{equation}
A_p=\int_0^{2\pi}\left(J_{1_{up}}-J_{1_{lo}}\right)\ud\phi_1.
\label{eq:area_int}
\end{equation}
In \eqref{eq:area_int}, each point of $J_{1_{up}}$ and $J_{1_{lo}}$ is also calculated numerically through the implicit functions $J_{1_{up}}(\phi_1)$ and $J_{1_{lo}}(\phi_1)$ mentioned above. The obtained results are reported in \autoref{fig:area_h_0_7475_nt2} by a solid line. For intermediate values of $L$, just after the maximum of the curve $A_p=A_p(L)$, there is an ambiguity concerning whether the boundary is determined by the constraint $R_3=0$ or $R_2=0$, because for some values of $\phi_1$ the boundary is defined by the former, while for others it is defined by the latter relation. In this region we cannot calculate the size of the permitted area by \eqref{eq:area_int} and the calculation from the PSSs is more reliable. In this case we estimate the size of the permitted area as the sum of the areas of all the small rectangles of the grid on the PSS, which is also used in the scan maps, and correspond to permitted orbits. The obtained results  are depicted by dots in \autoref{fig:area_h_0_7475_nt2}. The two well computed by \eqref{eq:area_int} parts of $A_p(L)$ are connected in this region by a dashed straight line in order to obtain a continuous curve. It is worth noting that even this rough approximation is in good agreement with the results obtained by counting the permitted initial conditions on the PSS. The good agreement of the results obtained by these two approaches indicates that the used grid of initial conditions is satisfactorily dense for capturing the dynamics of the system.

\subsubsection{Regular and Chaotic configurations}

\begin{figure}[ht]
	\centering
	\begin{subfigure}[b]{0.3\textwidth}
	\caption{$h=-0.7475$, $L=-0.05$}
	\label{fig:01}
	\includegraphics[width=\textwidth]{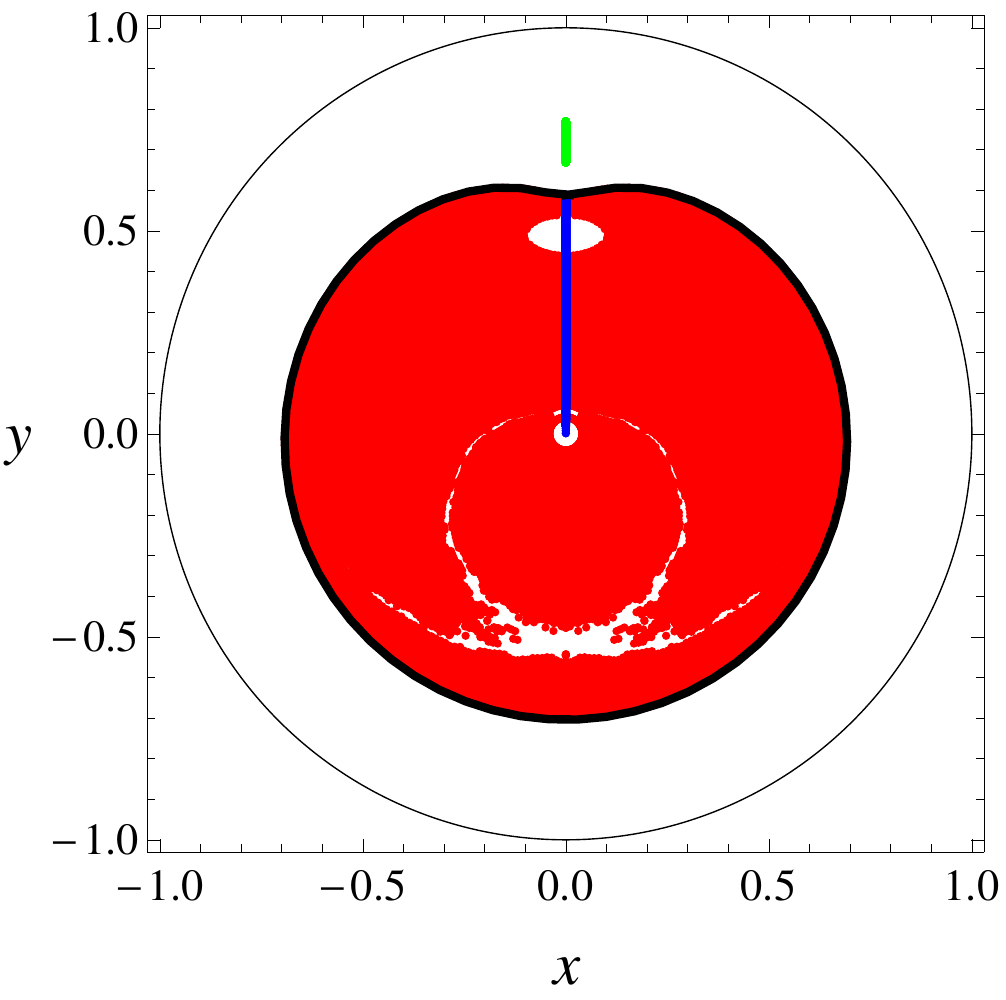}
	\end{subfigure}
	\begin{subfigure}[b]{0.3\textwidth}
	\caption{$h=-0.7475$, $L=0.05$}
	\label{fig:005}
	\includegraphics[width=\textwidth]{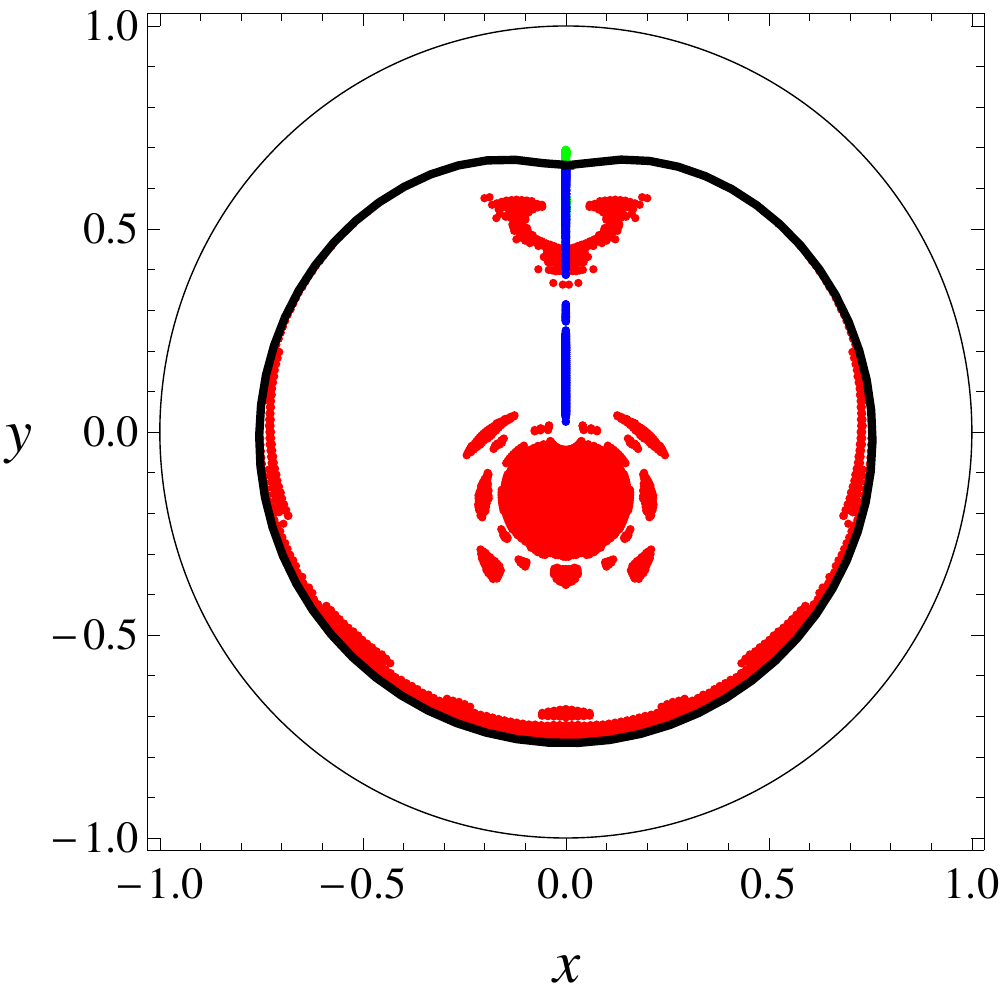}
	\end{subfigure}
	\begin{subfigure}[b]{0.3\textwidth}
	\caption{$h=-0.7475$, $L=0.25$}
	\label{fig:025}
	\includegraphics[width=\textwidth]{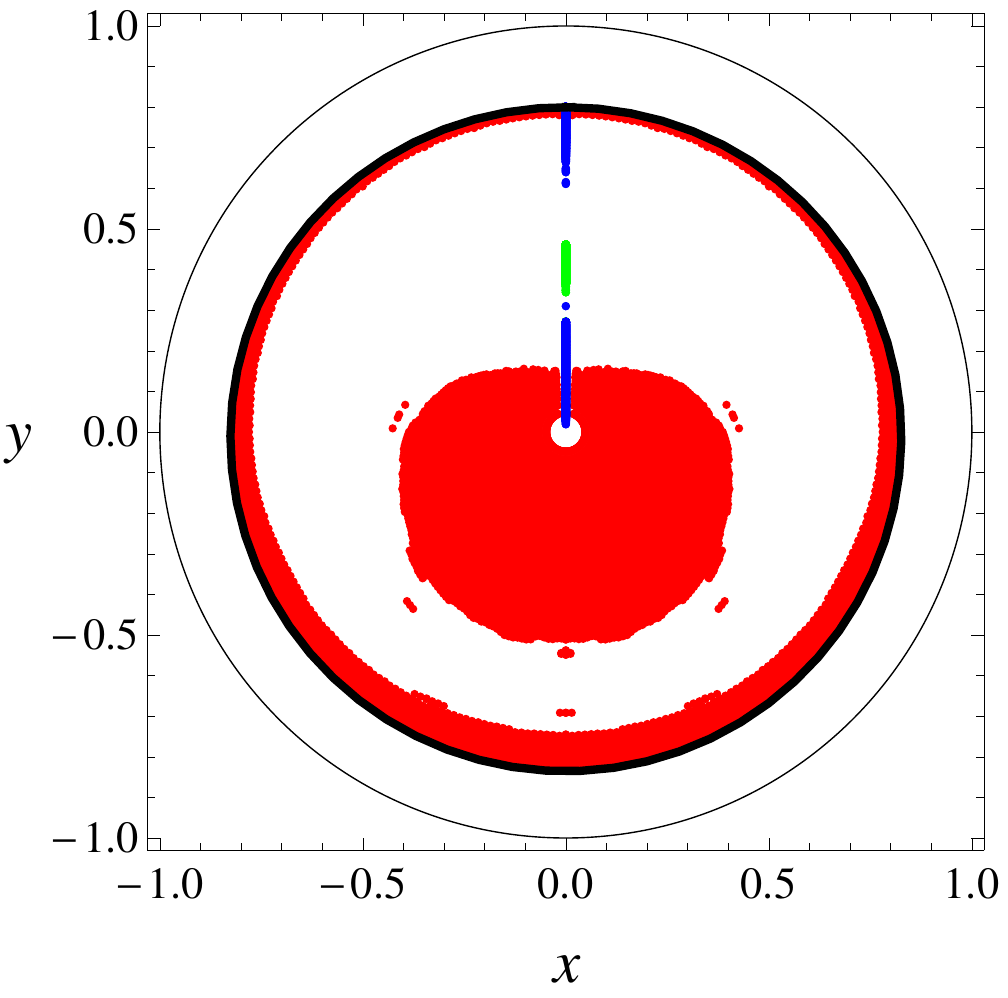}
	\end{subfigure}
	\begin{subfigure}[b]{0.3\textwidth}
	\caption{$h=-0.7475$, $L=0.35$}
	\label{fig:035}	
	\includegraphics[width=\textwidth]{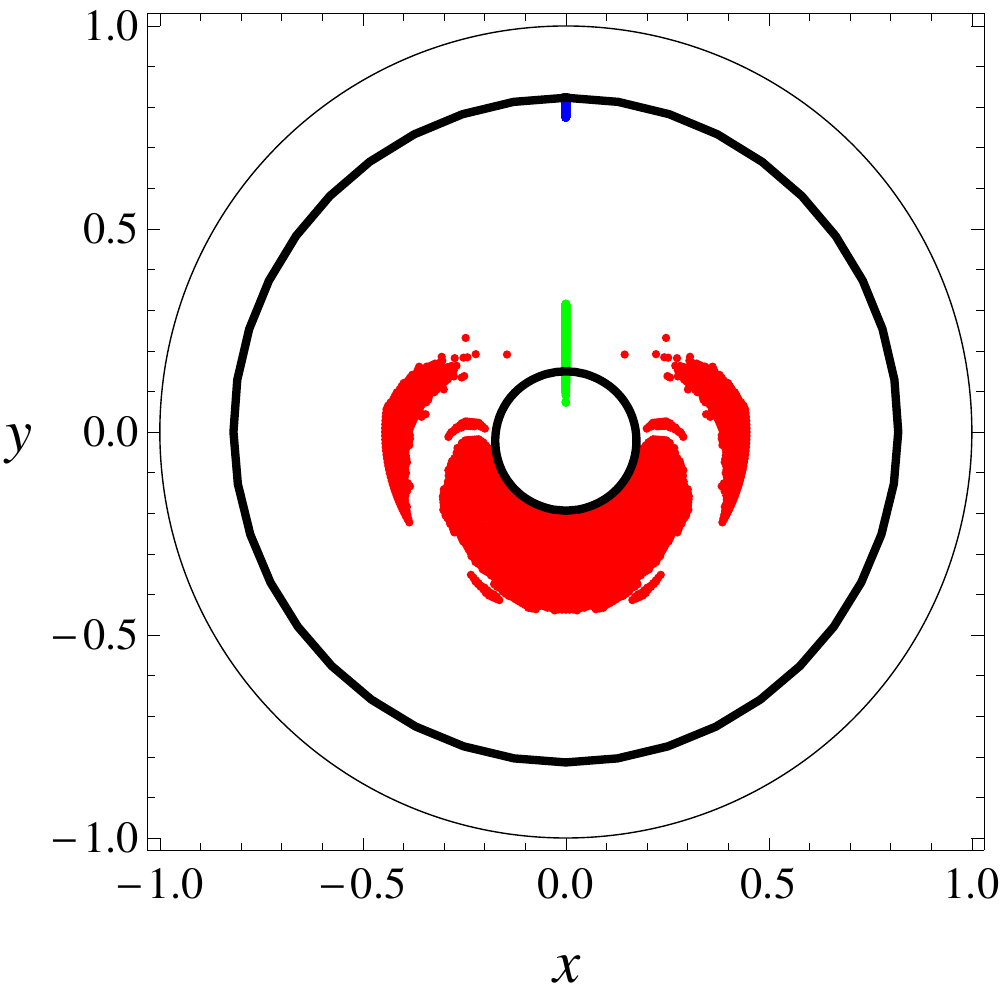}
	\end{subfigure}
	\begin{subfigure}[b]{0.3\textwidth}
	\caption{$h=-0.7475$, $L=0.45$}
	\label{fig:045}
	\includegraphics[width=\textwidth]{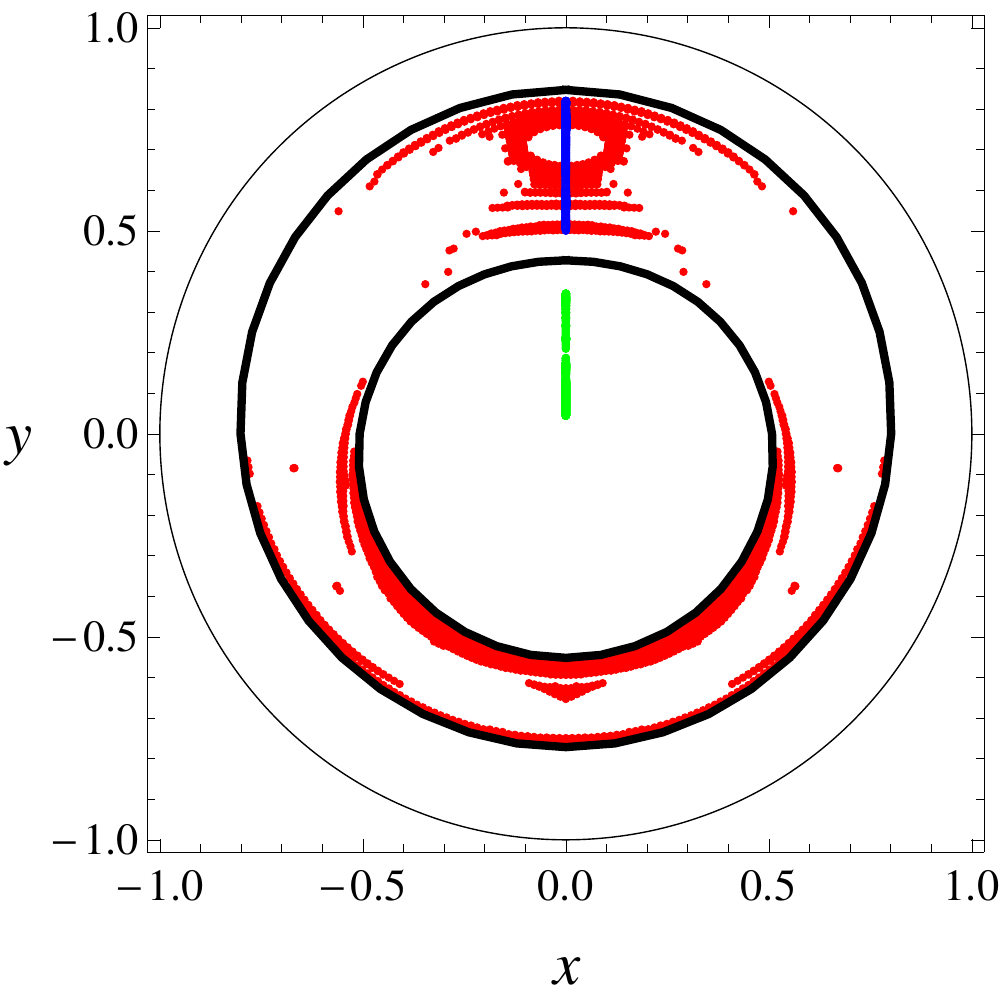}
	\end{subfigure}	
		\begin{subfigure}[b]{0.3\textwidth}
	\caption{$h=-0.7475$, $L=0.5$}
	\label{fig:055}
	\includegraphics[width=\textwidth]{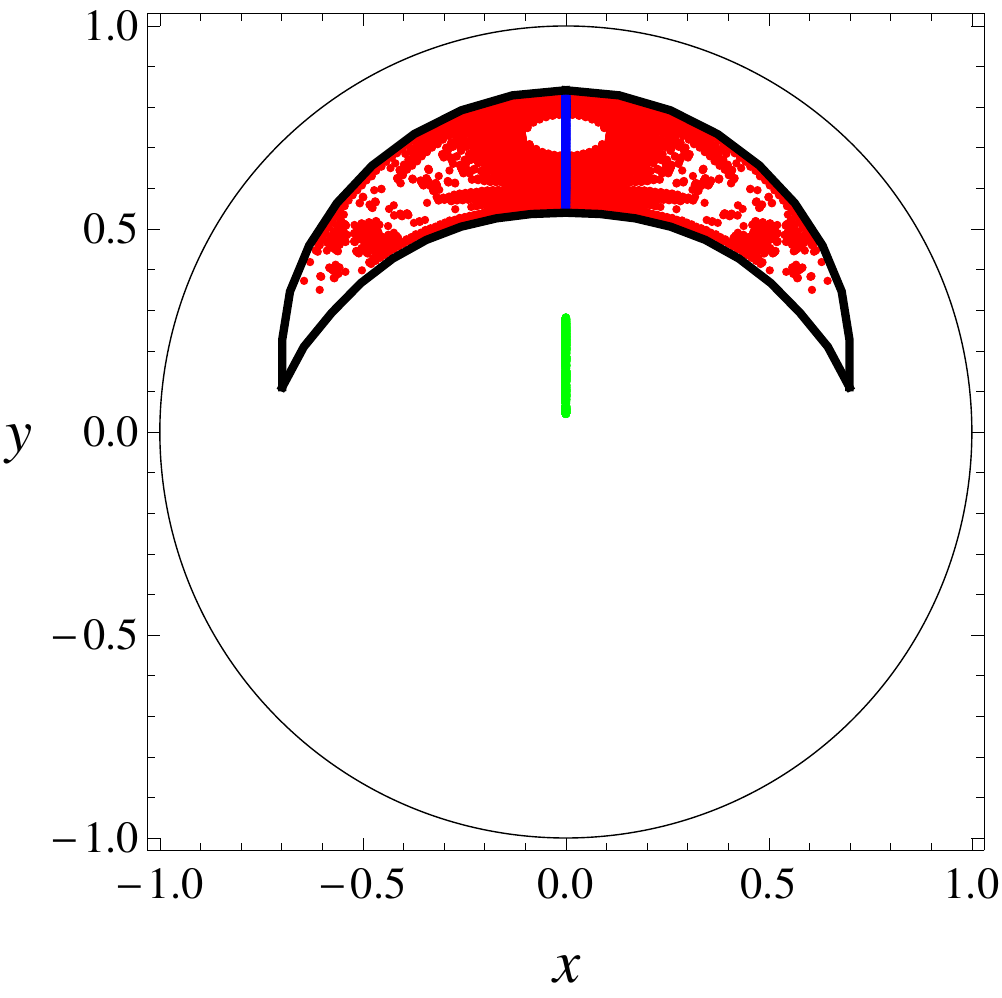}
	\end{subfigure}
\caption{Initial configurations which result to regular motion for $h=-0.7475$ and various values of $L$. The initial positions of $S_1$, $S_2$ and $S_3$ vortices are represented by red, green and blue color respectively. Since the configurations correspond to orbits on the PSS ($\phi_2=\pi/2$) the $S_2$ and $S_3$ vortices lie on the same semi-axis and some of the green dots may be indistinguishable. The permitted area of motion of $S_1$ is depicted by a thick black line. It can be easily seen that it is smaller than the disk defined by the Thomas-Fermi radius $R_{TF}=1$.}
\label{fig:confs}
\end{figure}

The chaotic or regular behavior of an orbit depends on the configuration (initial position in the $x-y$ plane of the BEC) of the vortices. In \autoref{fig:confs} the configurations which correspond to orbits on the PSS which exhibit regular motion are shown. Since these configurations correspond to initial conditions with $\phi_2=\pi/2$ the $S_2$ and $S_3$ vortices lie on the same half-line, while the $S_1$ vortex can occupy various positions. The initial positions of the $S_1$, $S_2$ and $S_3$ vortices are depicted in these figures by red, green and blue color respectively. The permitted area of  motion of the $S_1$ vortex is defined  by thick black lines.  In general the vortices in the BEC can move up to the Thomas-Fermi radius, which is equal to $R_{TF}=1$, but since we have fixed values of $h$ and $L$ the actual permitted area is smaller.

In \autoref{fig:confs} (in direct comparison also with \autoref{fig: sections_scans}) it is shown that for small values of $L$ almost all of the permitted area is occupied by regular orbits. As $L$ increases, almost all of the available configurations are chaotic, for $L=0.25$ the percentage of the chaotic orbits presents a local minimum and for values of $L>0.45$ the permitted area shrinks significantly and almost all the orbits become regular.

Let us look a bit closer at the relation  between the initial configuration of the vortices and the system's dynamical behavior by studying in more detail three representative cases. 

\begin{figure}[htbp]
	\centering
	\begin{tabular}{cc}
	\qquad(a)&\qquad(b)\\
		 \includegraphics[width=6cm]{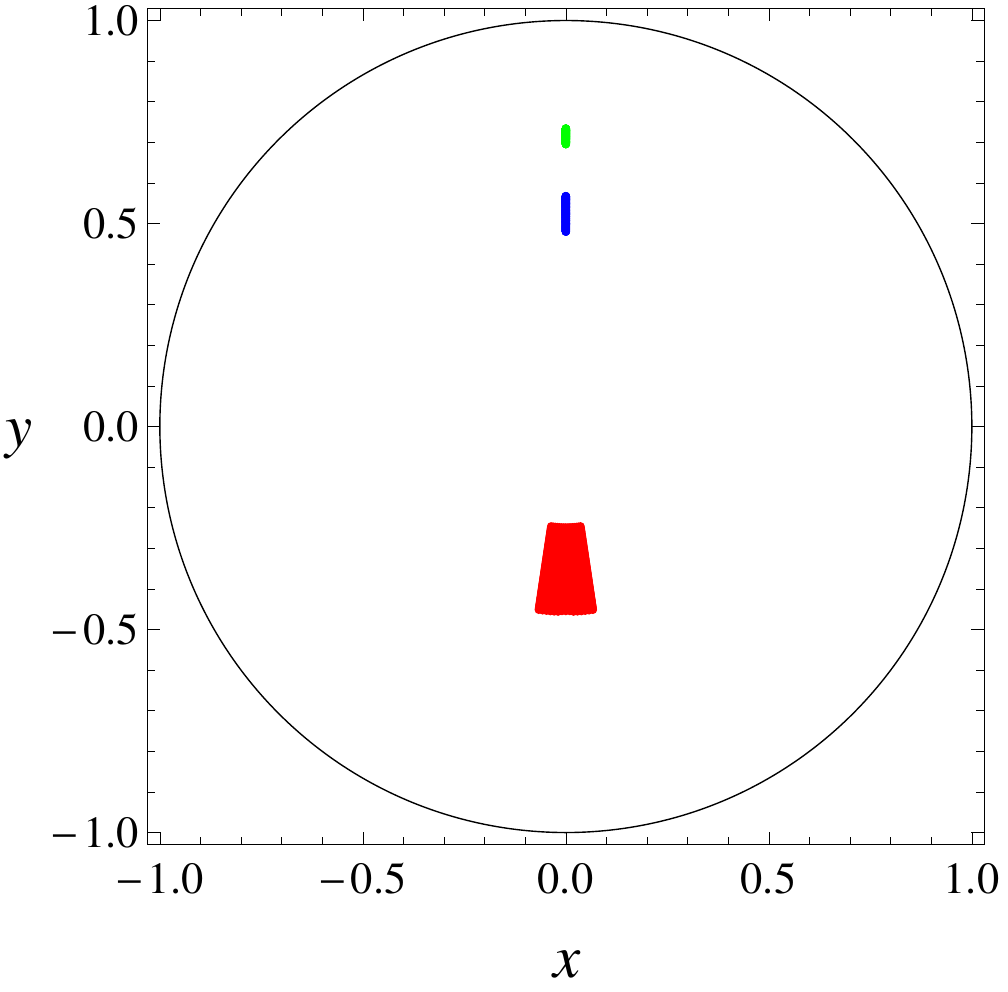}&\includegraphics[width=6cm]{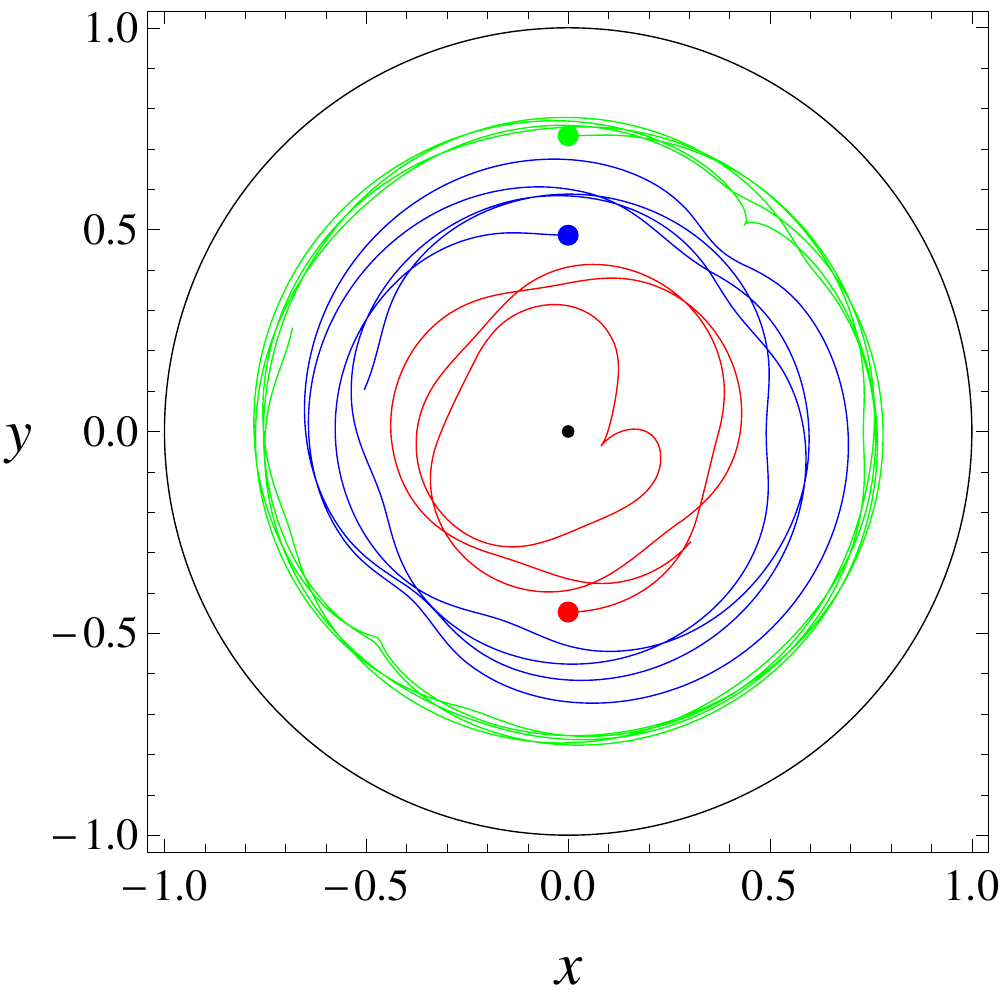}
		\end{tabular}
		\caption{(a) A fraction of the regular initial configurations in the $L=-0.05$ case which corresponds to the `rotational' regime. (b) The time evolution of  a representative orbit of this regime with $\phi_1=\pi$ and $J_1$=0.1. We can see that the trajectories do not intersect each other, so the dynamics belongs to the `one--vortex' regime.}
	\label{fig:conf_m005_part1}
\end{figure}

We start our analysis by considering $L=-0.05$ for which almost all initial vortex configurations (or initial conditions) lead to regular motion (see \autoref{fig: sections_scans}(b) and \autoref{fig:confs}(a)).  In \autoref{fig:conf_m005_part1}(a) we consider an ensemble  of initial configurations in order to check the motion corresponding to it. In \autoref{fig:conf_m005_part1}(b) the time evolution of a representative orbit with $\phi_1=\pi$ and $J_1=0.1$ is shown. The initial condition of this orbit corresponds to point A in the PSS of \autoref{fig: sections_scans}(b). This is the case of the `rotational' regime where all the vortices rotate around the common center of rotation without any major disturbances to their motion, producing regular behaviors. As we can see, the areas the orbits of the individual vortices occupy are distinct and  do not mix. In this case the vortices interact weakly with each other and they are said to be at  the so-called `one--vortex' regime. In general, when an initial configuration produces evolutions belonging to the one-vortex regime the resulting motion is regular.

\begin{figure}[htbp]
	\centering
		\begin{tabular}{cc}
	\qquad(a)&\qquad(b)\\
	 \includegraphics[width=6cm]{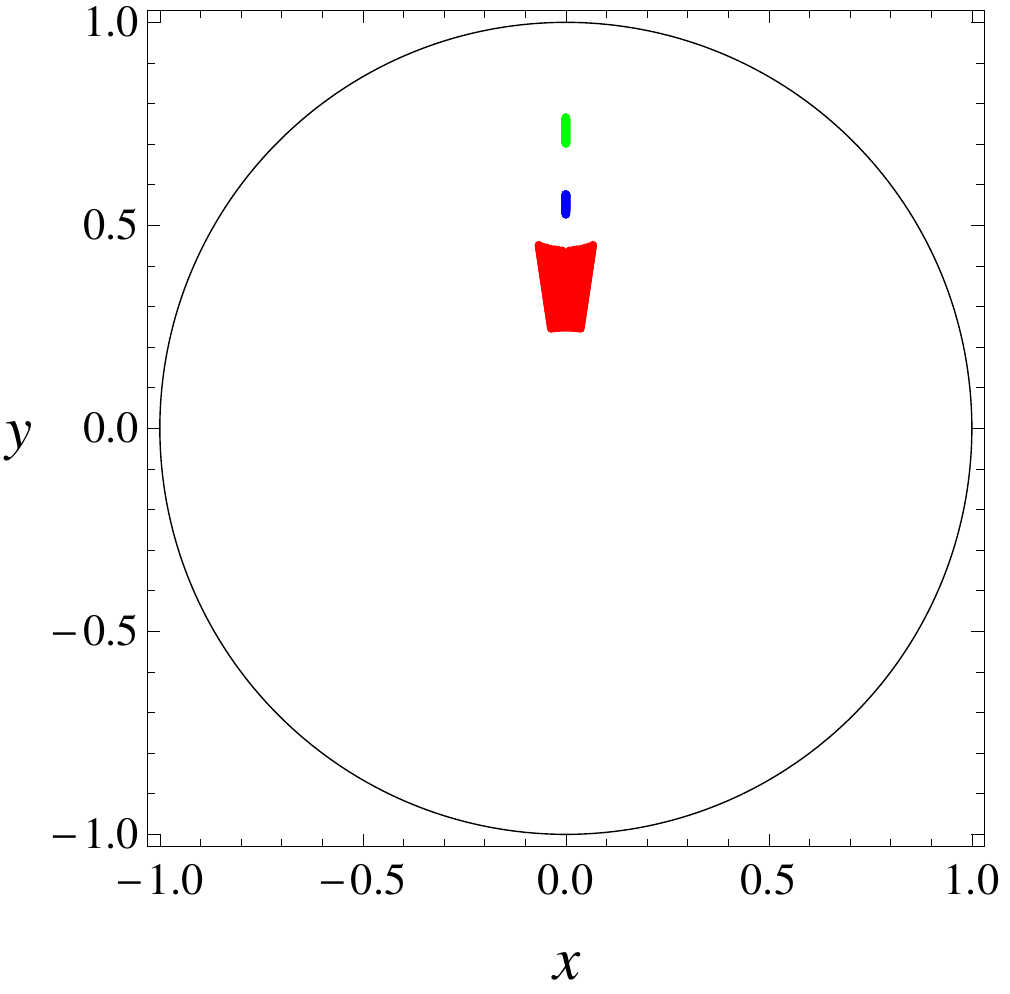}&\includegraphics[width=6cm]{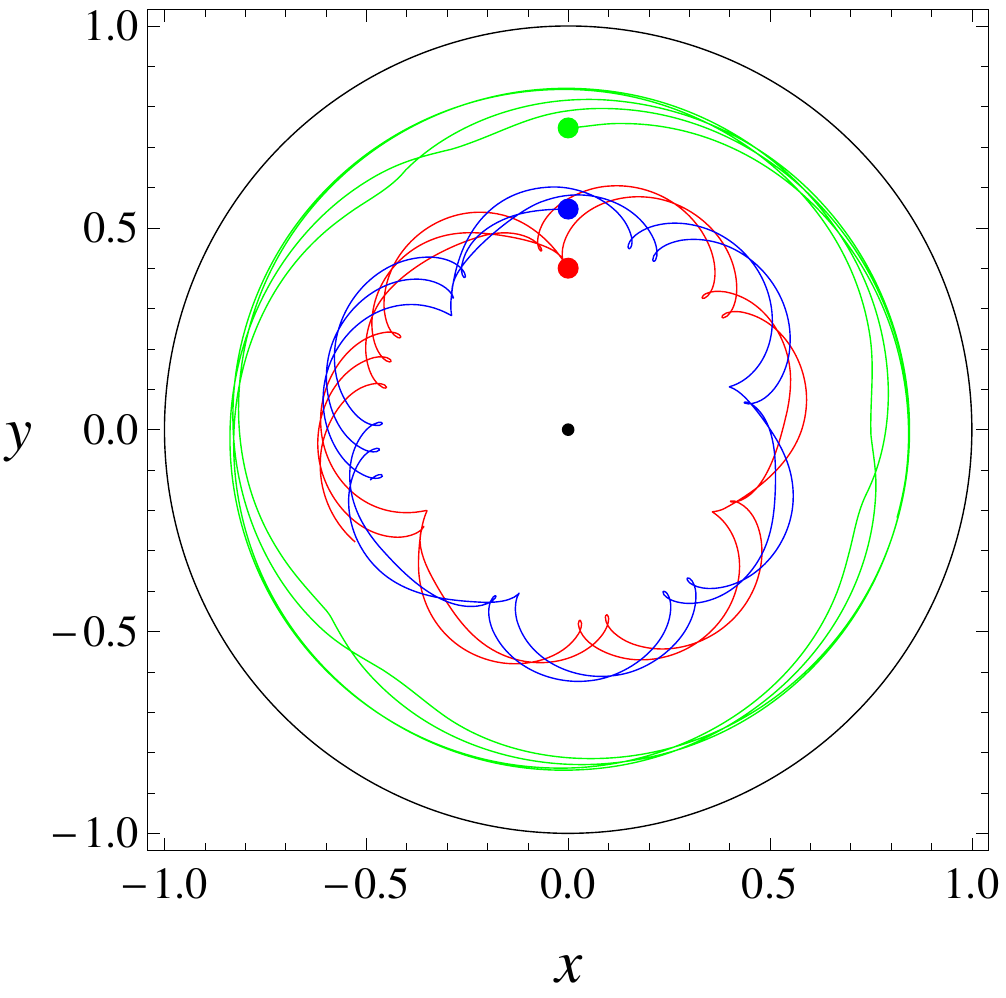}
	\end{tabular}
		\caption{(a) A fraction of the regular initial configurations in the $L=-0.05$ case which corresponds to the `satellite' regime. (b) The time evolution of  a representative orbit of this regime with $\phi_1=0$ and $J_1$=0.08. We can see that $S_1$ and $S_3$ interact strongly with each other but their orbits do not intersect that of $S_2$, so we are inside the `two--vortex' regime.}
		\label{fig:conf_m005_part2}
\end{figure}

The second case we examine is shown in \autoref{fig:conf_m005_part2}. In this case the $S_1$ and $S_3$ vortices rotate around each other and both of them around the center of rotation. The $S_2$ vortex rotates around the center as well but in the opposite direction because of its opposite charge. This is the so-called `satellite' regime. In this case the two vortices interact more strongly with each other while they exhibit a  weak interaction with the third one. As we can see in \autoref{fig:conf_m005_part2}(b)  the areas of  trajectories of $S_1$ and $S_3$ cover overlap but do not mix with the one of $S_2$. This dynamical regime is referred as the  `two--vortex' regime which also results in regular motion. The corresponding initial conditions of this orbit are depicted in the PSS of \autoref{fig: sections_scans}(b) by the point B.

\begin{figure}[htbp]
	\centering
	\begin{tabular}{ccc}
	\qquad(a)&\qquad(b)&\qquad(c)\\		
	 \includegraphics[width=5cm]{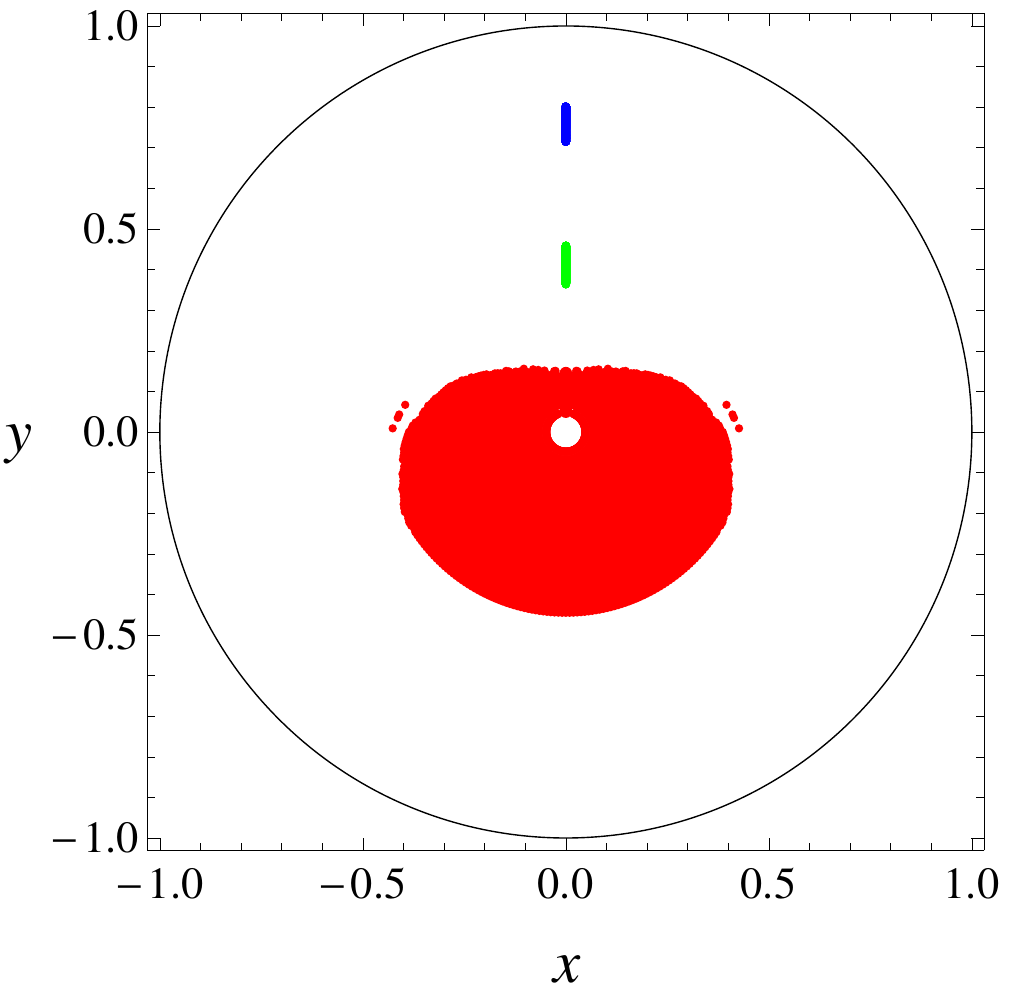}&\includegraphics[width=5cm]{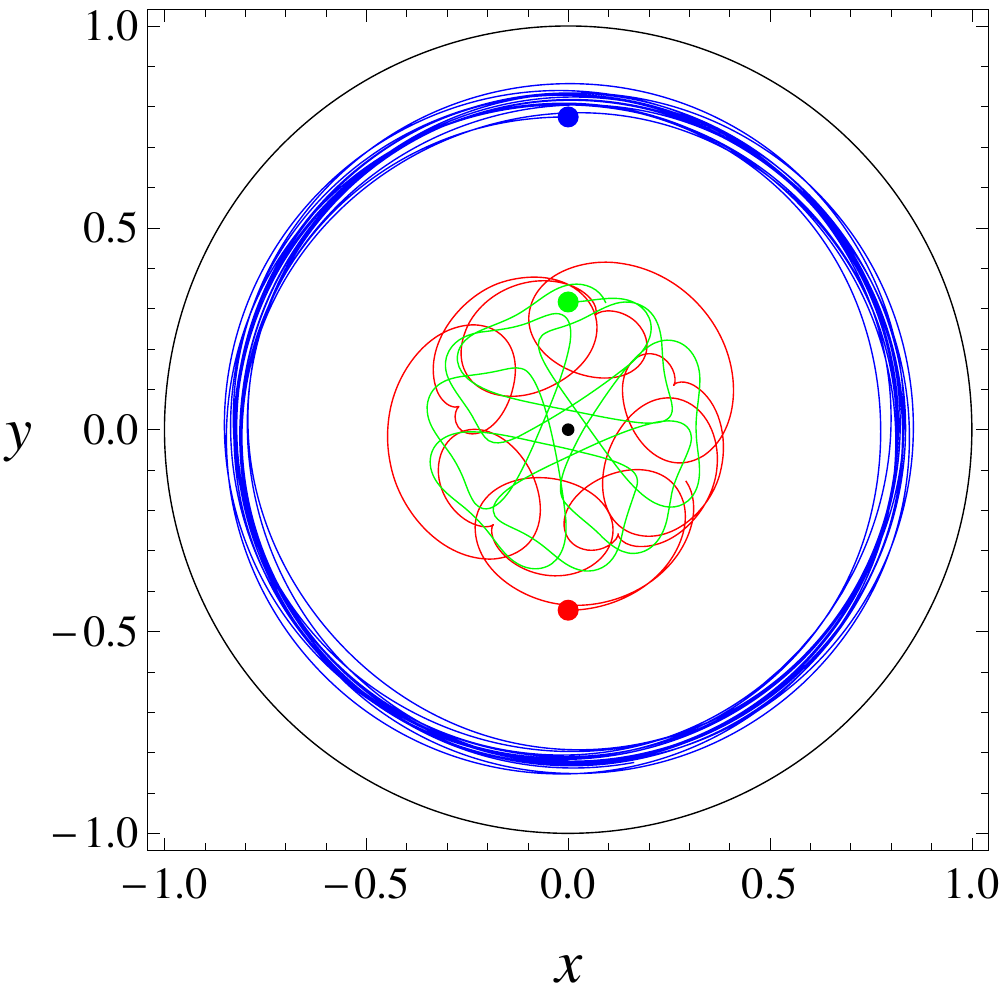}&\includegraphics[width=5cm]{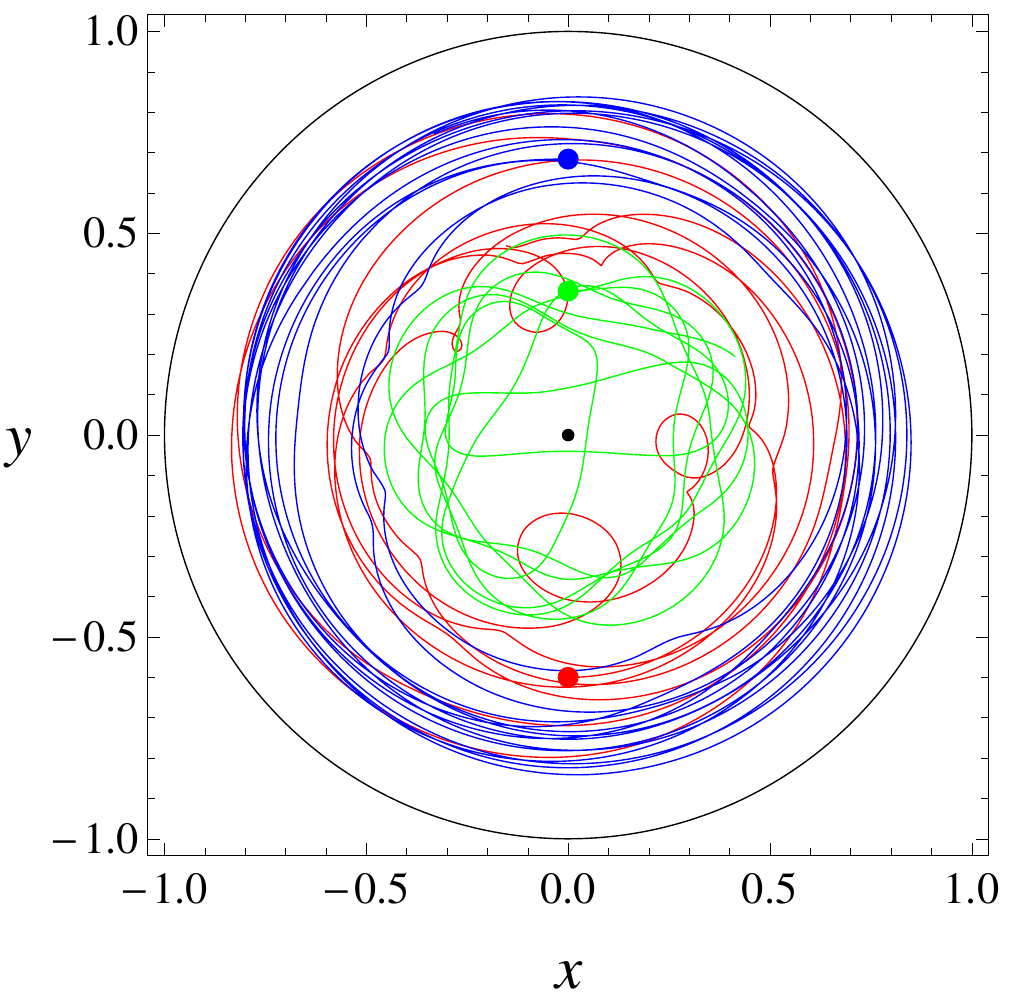}
	\end{tabular}
	\caption{(a) A fraction of the regular initial configurations in the $L=0.25$ case which corresponds to $J_1\lesq 0.1$. (b) The time evolution of  a representative orbit of this regime with $\phi_1=\pi$ and $J_1=0.1$, which corresponds to a `two--vortex' configuration. (c) The  time evolution of the chaotic orbit with $\phi_1=\pi$ and $J_1=0.18$. The trajectories of all  vortices are mixing with each other, leading to chaotic behavior.}
	\label{fig:conf_35_part}
\end{figure}

The next case we consider is  the one with $L=0.25$. For this value of $L$  
numerous chaotic as well as regular initial configurations exist. In \autoref{fig:conf_35_part}(a) the regular initial configurations for $J\lesq0.1$ are depicted. In \autoref{fig:conf_35_part}(b) the time evolution of a representative orbit of this ensemble with $\phi_1=\pi$ and $J_1=0.1$ is depicted. The initial condition of this orbit is depicted in the upper panel of \autoref{fig: sections_scans}(d) as point C. The trajectories of the vortices clearly correspond to the `two--vortex' regime since only  two them interact strongly. On the other hand in  \autoref{fig:conf_35_part}(c) the time evolution of an initial configuration with $\phi_1=\pi$ and $J_1=0.18$ leading to chaotic motion is shown. The initial condition is shown in \autoref{fig: sections_scans}(d) as point D. Here the strong interaction between all the vortices, which is generally necessary in order to have chaotic motion, can be concluded by the fact that the orbits of all the vortices mix with each other.

\subsubsection{Consideration of an alternative $t_{max}$ based on the physical aspects of the system}

Let us now study in more detail the system's chaotic behavior. Since the physical model from which this study has been motivated is a Bose-Einstein condensate, which has a limited life time (commonly of
the order of a few seconds to a few tens of seconds),
there are some associated considerations to be kept in mind. In particular, in our set up the condensate's life time is of the order of a few hundreds up to one thousand time units.

\begin{figure}[h]
\begin{tabular}[t]{ccc}
	 \includegraphics[height=7cm]{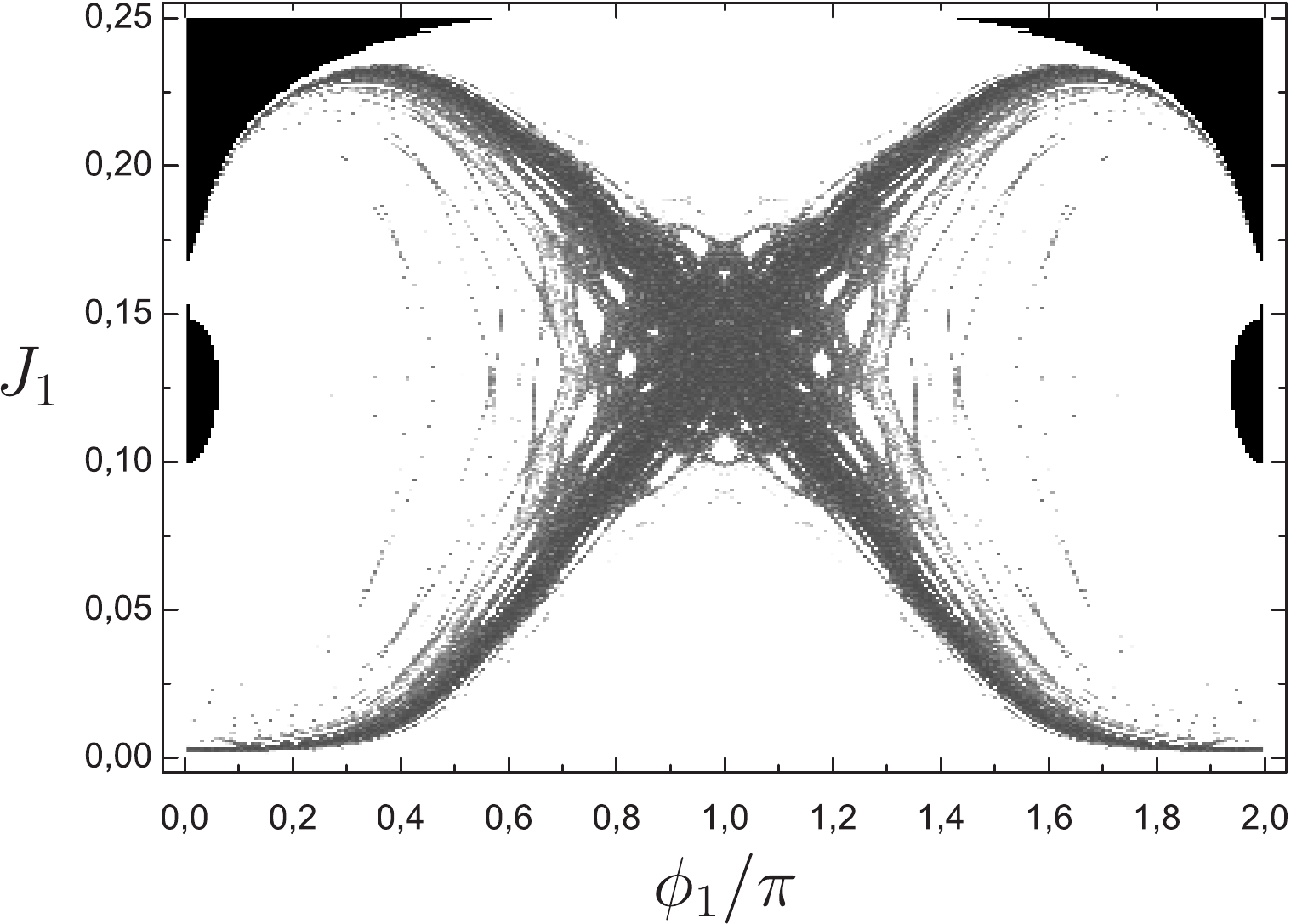}&$\quad$&\includegraphics[height=7cm]{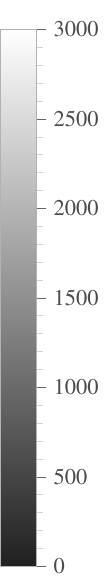}
\end{tabular}
\caption{Scan map for $h=-0.7475$ and $L=-0.03$. Dark colored points correspond to orbits with small $t_{S_0}$ (chaotic) and light colored points correspond to orbits with large $t_{S_0}$ (regular). This correspondence is also shown in the given legend.}
\label{fig: 003scan}
\end{figure}

\begin{figure}[h]
	\centering
	\begin{subfigure}[b]{0.4\textwidth}
	\caption{$140\lesq{}t_{S_0}\lesq500$}
	\label{fig: weak-strong-a}
	 \includegraphics[width=\textwidth]{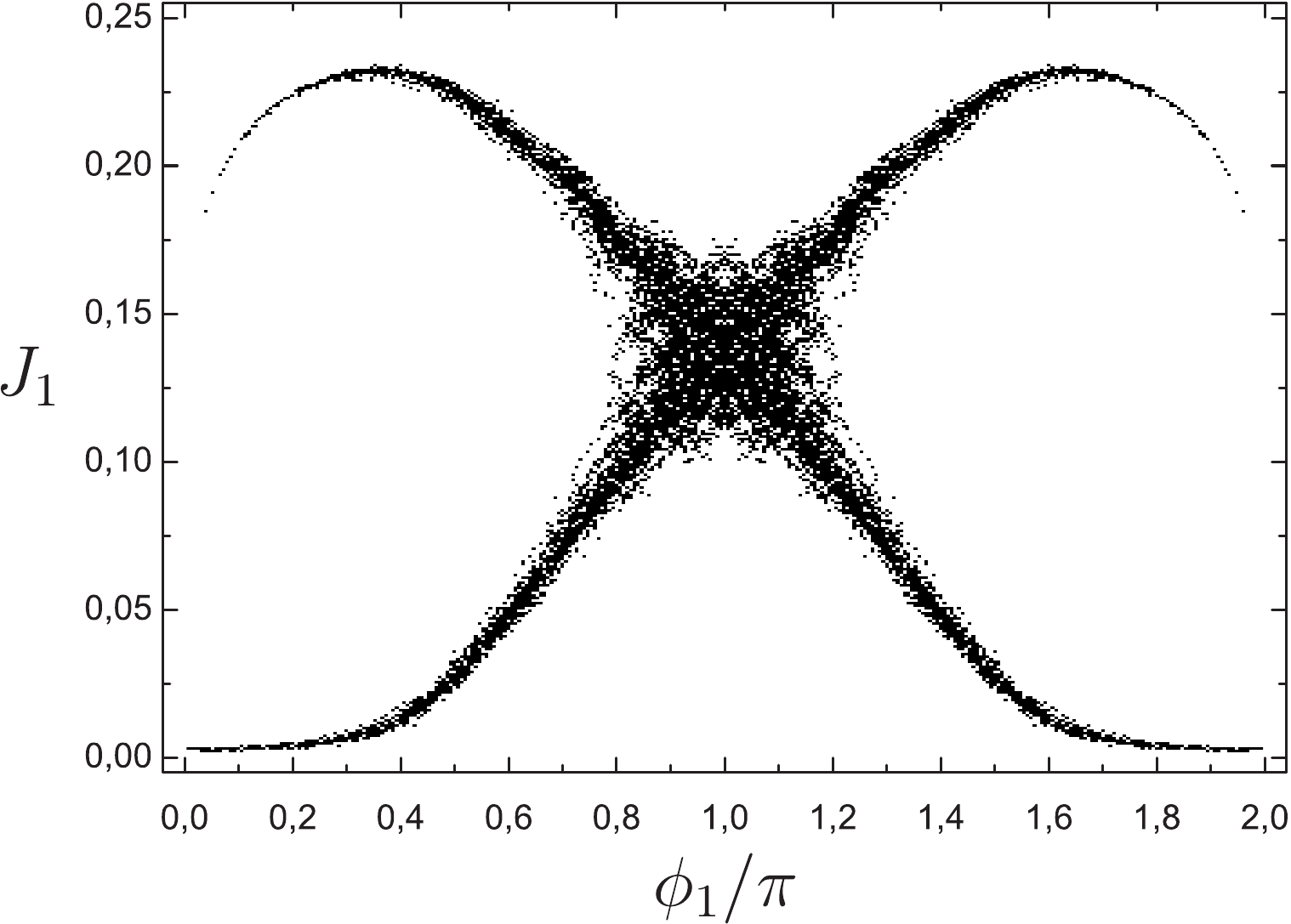}
	\end{subfigure}
	\begin{subfigure}[b]{0.4\textwidth}
	\caption{$500<t_{S_0}\lesq1000$}
	\label{fig: weak-strong-b}
	\includegraphics[width=\textwidth]
	{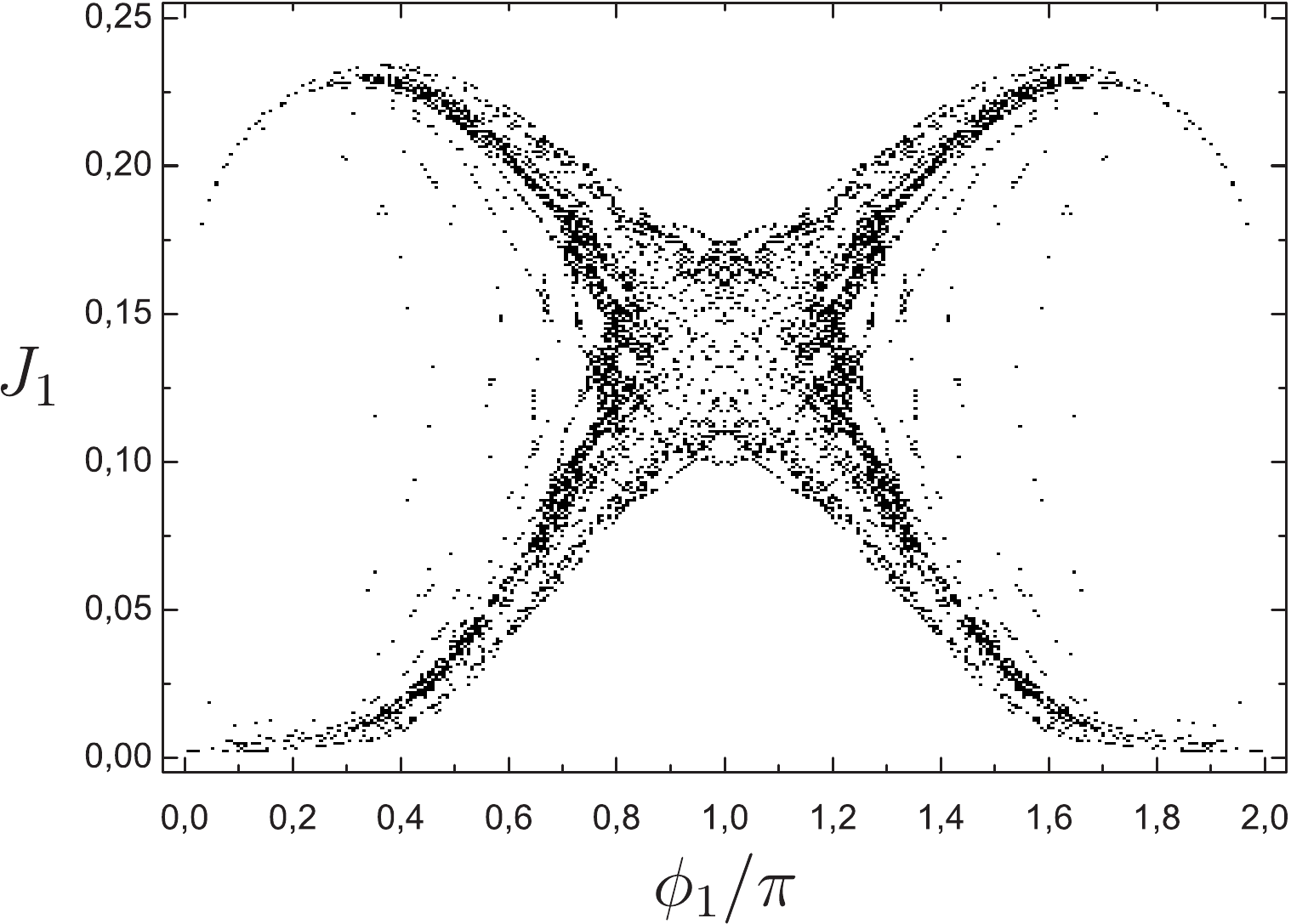}
	\end{subfigure}
	\begin{subfigure}[b]{0.4\textwidth}
	\caption{$1000<t_{S_0}\lesq1500$}
	\label{fig: weak-strong-c}
	 \includegraphics[width=\textwidth]{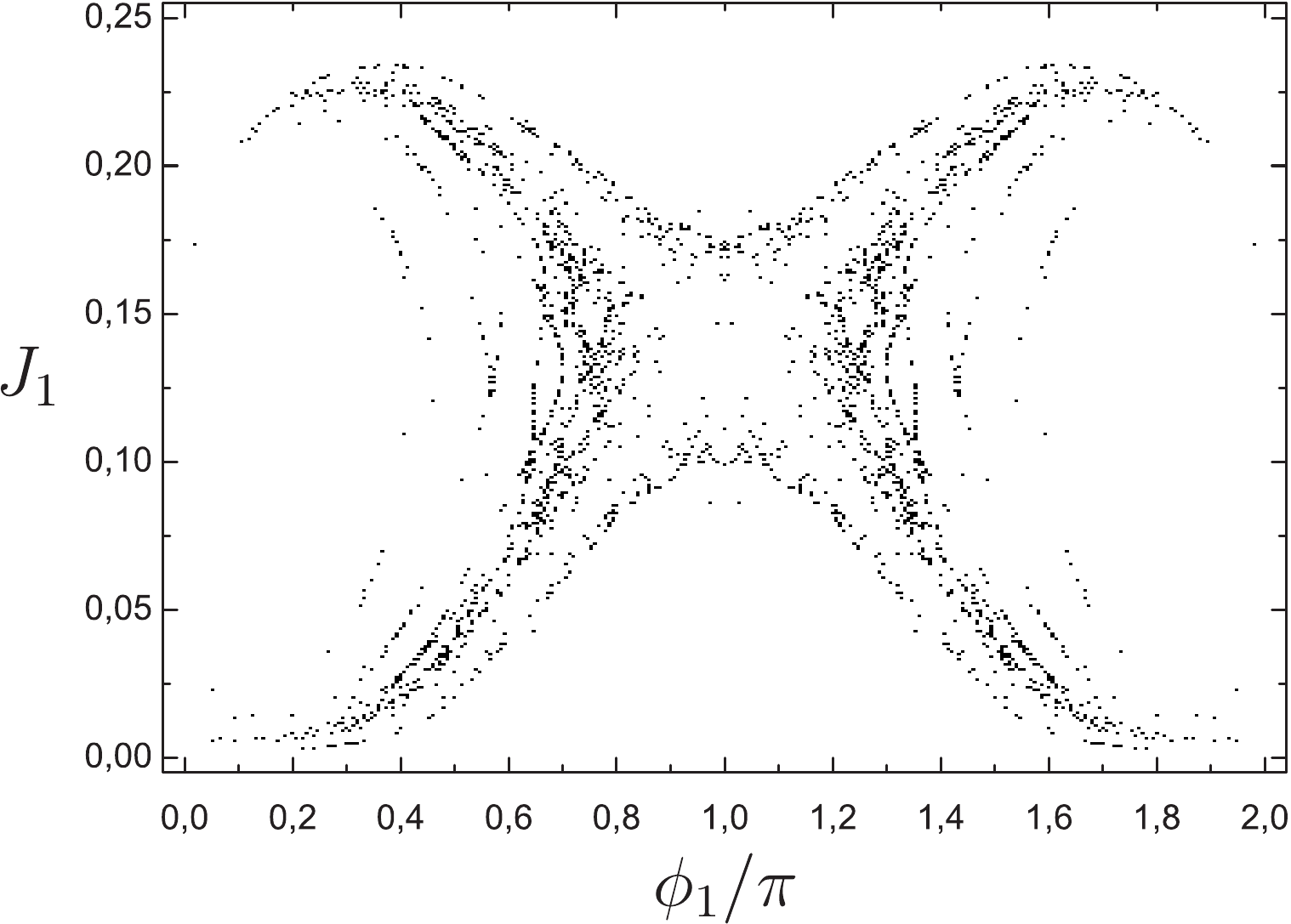}
	\end{subfigure}
	\begin{subfigure}[b]{0.4\textwidth}
	\caption{$1500<t_{S_0}\lesq2000$}
	\label{fig: weak-strong-d}
	 \includegraphics[width=\textwidth]{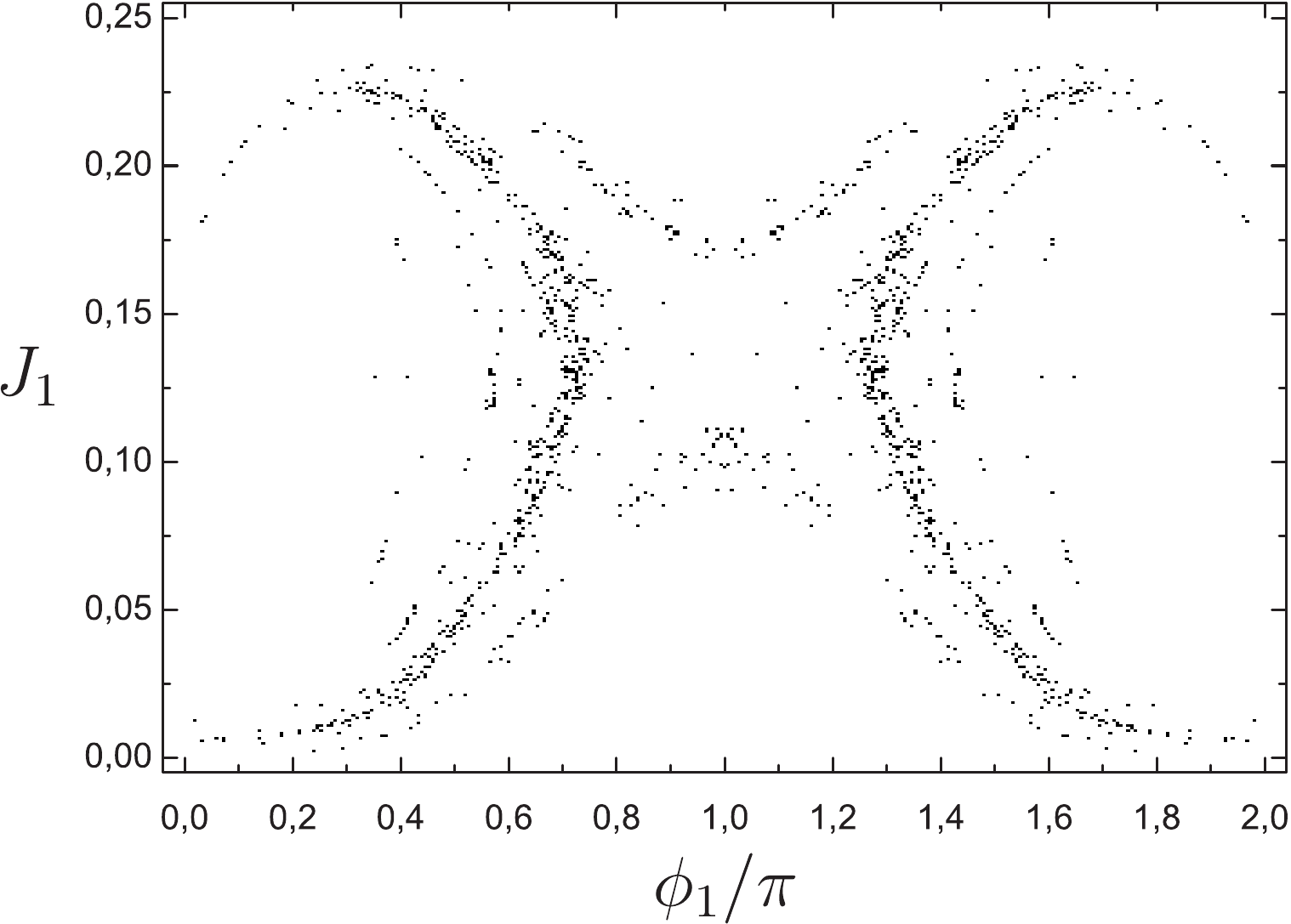}
	\end{subfigure}	
\caption{The decomposition of \autoref{fig: 003scan} for various intervals of $t_{S_0}$. We see that as $t_{S_0}$ increases the corresponding points of the scan map lie further away from the center of the x-shaped chaotic region and their number decreases.}
\label{fig: weak-strong}
\end{figure}

We thus need to explore the implications that this physically induced time limit has. In order to address this question we consider in \autoref{fig: 003scan} the scan map for $h=-0.7475$ and $L=-0.03$. This map was constructed similarly to the ones of \autoref{fig: sections_scans}. The connected chaotic region in the center of this plot (dark gray points) can be constructed by any orbit starting in it. Nevertheless, depending on where we choose the initial condition of this orbit, the time in which its chaotic behavior is revealed varies. This becomes evident from the results of \autoref{fig: weak-strong}
where the plot of \autoref{fig: 003scan} is decomposed into four regions depending on the $t_{S_0}$ values of the initial conditions. In particular we consider points with  $t_{S_0}\in[140,500]$ (\autoref{fig: weak-strong-a}),
$t_{S_0}\in(500,1000]$ (\autoref{fig: weak-strong-b}), $t_{S_0}\in(1000,1500]$ (\autoref{fig: weak-strong-c}) and $t_{S_0}\in(1500,2000]$ (\autoref{fig: weak-strong-d}). From these figures we can conclude that as we move further from the center of the x-shaped region the orbits become `stickier' and thus they require more time to reveal their chaotic nature. These orbits  involve predominantly two-vortex dynamics during earlier stages of the evolution, while at later stages all three vortices are interacting with each other, leading to the associated observed chaotic features. Since the typical lifetime of the BEC is a few hundred time units, a good candidate for a physically meaningful integration time would be $t_{max}=500$. In this way, chaotic orbits which reveal their nature later than this time can be considered, from a practical point of view, as regular. For instance, in real experiments one would expect to detect chaotic motion in the limited time that the experiment lasts, only in regions with small $t_{S_0}$. In our case, such orbits are the ones plotted in \autoref{fig: weak-strong}(a), whose initial conditions are located close to the center of the x-shaped chaotic region. Thus, the need for efficient chaos indicators, capable of determining the nature
of the orbits in potentially shorter, physically meaningful time intervals is of considerable importance. The SALI can successfully play this role, since it can reveal the larger part of the chaotic region of the system, even for $t_{S_0}=500$, as can be seen in \autoref{fig: weak-strong} and will be also shown in the next section through \autoref{fig:pcnt_07475_both}. On the other hand the mLE would require at least an order of magnitude larger integration times in order to acquire decisive results, which is both physically irrelevant and CPU-time consuming.


In what follows we will both use $t_{max}=3000$ in order to reveal the full dynamics of the system and $t_{max}=500$ in order to determine its `practical' dynamical behavior. The different choices of $t_{max}$ will be clearly indicated.

\subsubsection{Chaoticity Percentages}

\begin{figure}
	\centering
		\includegraphics[width=8cm]{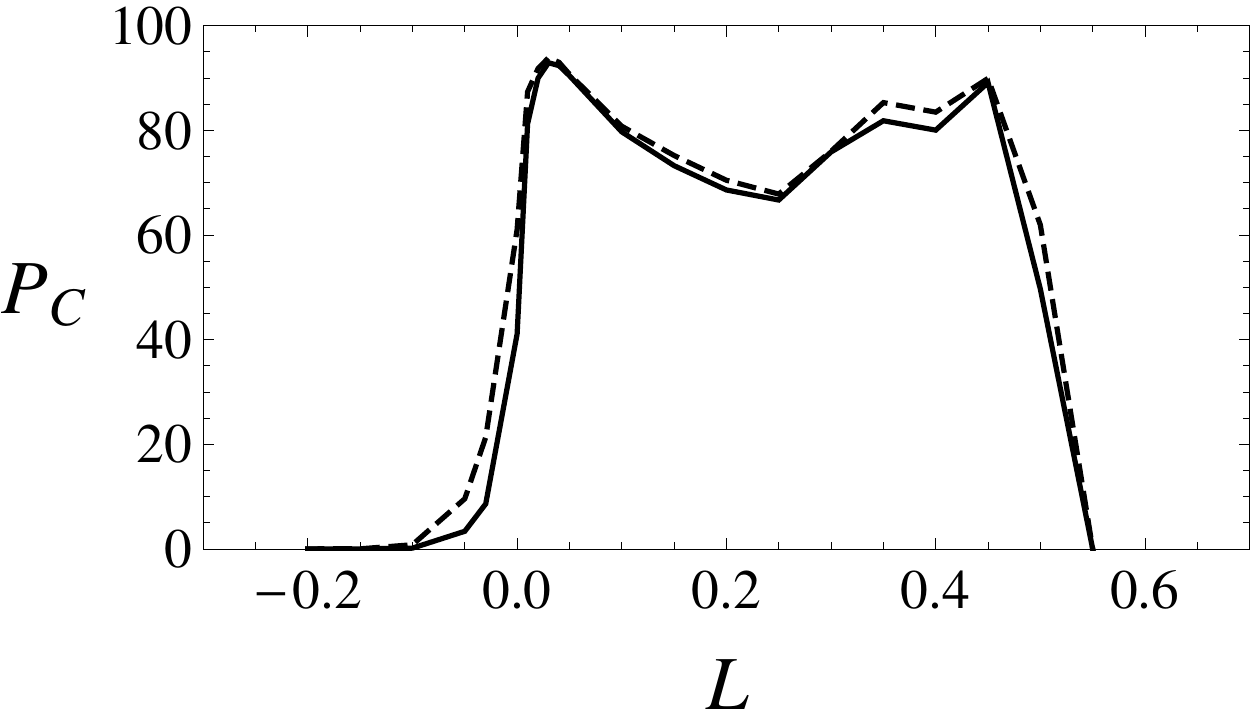}
		\caption{The percentage $P_C$ of initial conditions leading to chaotic orbits over the total number of permitted ones with respect to the value of $L$ for $h=-0.7475$. The solid line represents the results for $t_{max}=500$ while the dashed one corresponds to $t_{max}=3000$. We observe that the main characteristics of $P_C$ are the same in both cases.}
	\label{fig:pcnt_07475_both}
\end{figure}

In order to have a complete picture of the evolution of the chaotic region for varying $L$ we calculated, using the SALI, the percentage of the chaotic orbits over the permitted ones (\autoref{fig:pcnt_07475_both}). For this calculation we used both $t_{max}=500$ as well as $t_{max}=3000$. We can see that the percentage of the chaotic region is larger when $t_{max}$ is larger, since some of the sticky orbits are now characterized as chaotic, but the general behavior does not change significantly. 
By examining \autoref{fig:confs} the results of \autoref{fig:pcnt_07475_both} can be easily understood. As was previously explained, for small $L$ ($L\lesq-0.05$) the initial configurations correspond to either the one-- or the two--vortex regime, which leads to regular motion. As $L$ increases the orbit of either $S_1$ or $S_3$ lies further from the rotation center than before. There, it approaches the orbit of $S_2$, causing  strong interactions between all three vortices through slingshot effects, which result to a maximization of the chaotic region at $L=0.05$. As the value of $L$ increases further, one of the $S_1$, $S_3$ vortices can lie far enough from the other two in order for the system to have two--vortex configurations, causing in this way the local minimum of the chaotic percentage for $L=0.25$ observed in \autoref{fig:pcnt_07475_both}. At $L=0.35$ we obtain a secondary maximum of the chaoticity percentage. This maximum corresponds to the range of $L$ values just after the maximum of the permitted area (\autoref{fig:area_h_0_7475_nt2}), where the ambiguity in the calculation of $A_p$ occurs. This happens because at this $L$ value the upper boundary of the permitted area is defined by both the $R_1=0, R_3=0$ and $R_2=0$ constraints, which means that the orbit of either $S_1$ or $S_3$ is close to the one of $S_2$, causing again strong interactions between all three vortices and thus leading to chaotic behavior. For $L=0.45$ both vortices with positive charge lie at a large distance from the center (nevertheless comparable to the one of $S_2$) but, at the same time, the permitted area of motion has decreased, which allows mainly configurations with strong interactions between all three vortices. This behavior leads to the maximization of the chaotic region. Finally, for even larger values of $L$ ($L>0.45$) the permitted area of $S_1$ and $S_3$ becomes very narrow and is located far away from the $S_2$ vortex, leading to two-vortex configurations and consequently to the minimization and eventual disappearance of the chaotic region.

\subsection{Global Dynamics for $-1.1\lesq h\lesq -0.5$}

Based on the detailed analysis presented for the representative case of $h=-0.7475$ we perform a global study of the system's dynamical behavior for energy values in the range  $-1.1\lesq h\lesq -0.5$ in order to discover if the system exhibits similar behaviors. This range is assumed to contain all the physically meaningful energy values of the system.

\paragraph{Permitted area of motion.}

\begin{figure}[h]
	\centering
	\begin{subfigure}{0.3\textwidth}
	\centering
	\caption{$h=-1.1$}
	\includegraphics[width=\textwidth]{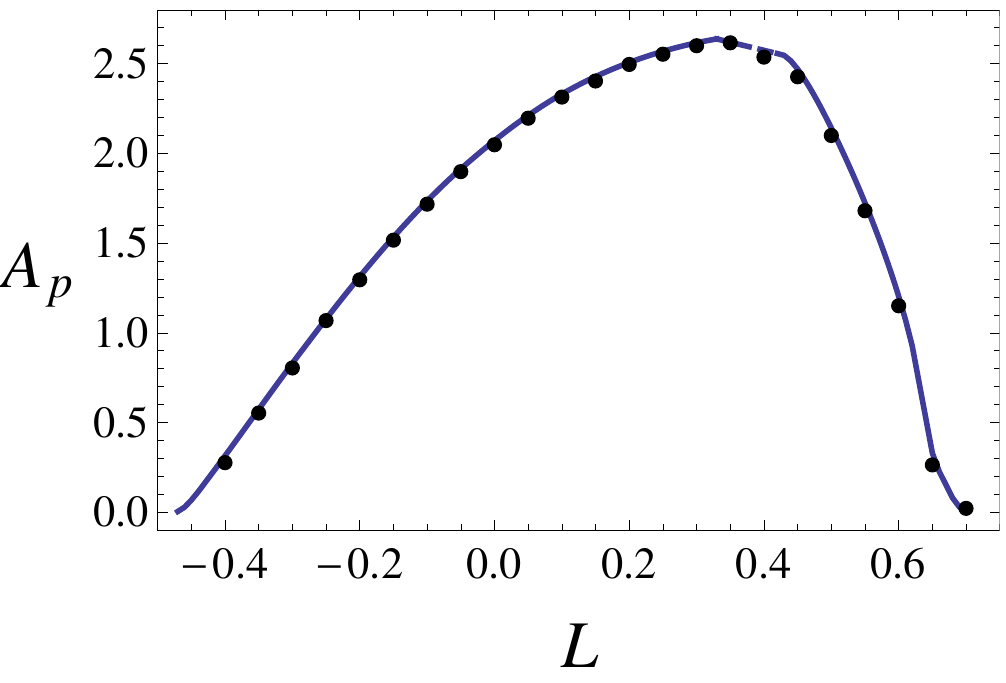}
	\label{fig: area_11}
	\end{subfigure}
	\begin{subfigure}{0.3\textwidth}
	\centering
	\caption{$h=-0.9$}
	\includegraphics[width=\textwidth]{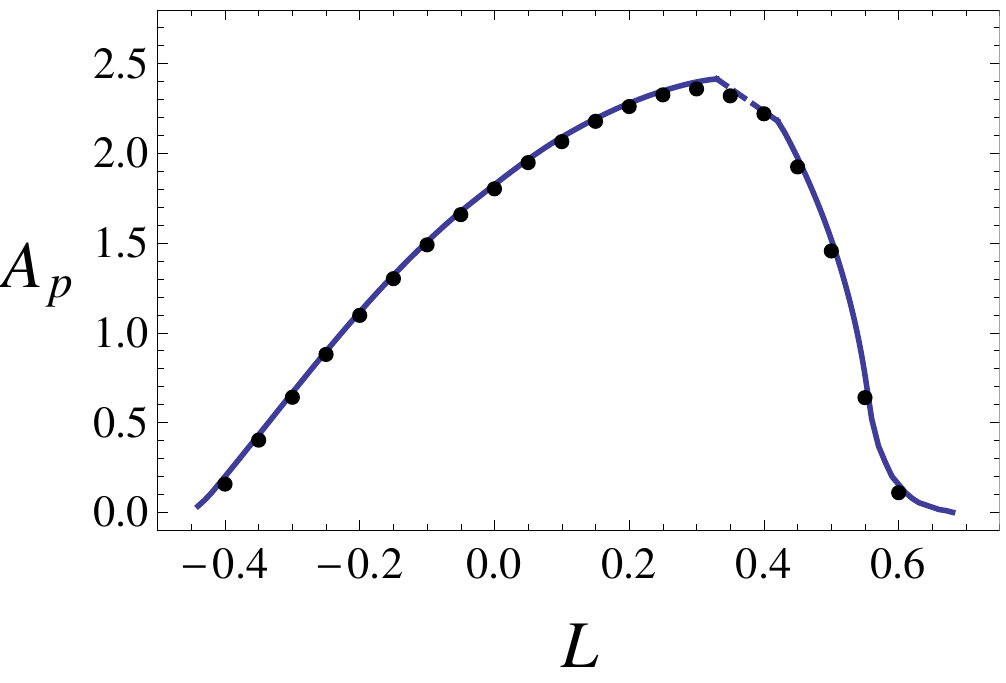}
	\label{fig: area_09}
	\end{subfigure}
	\begin{subfigure}{0.3\textwidth}
	\centering
	\caption{$h=-0.8$}
	\includegraphics[width=\textwidth]{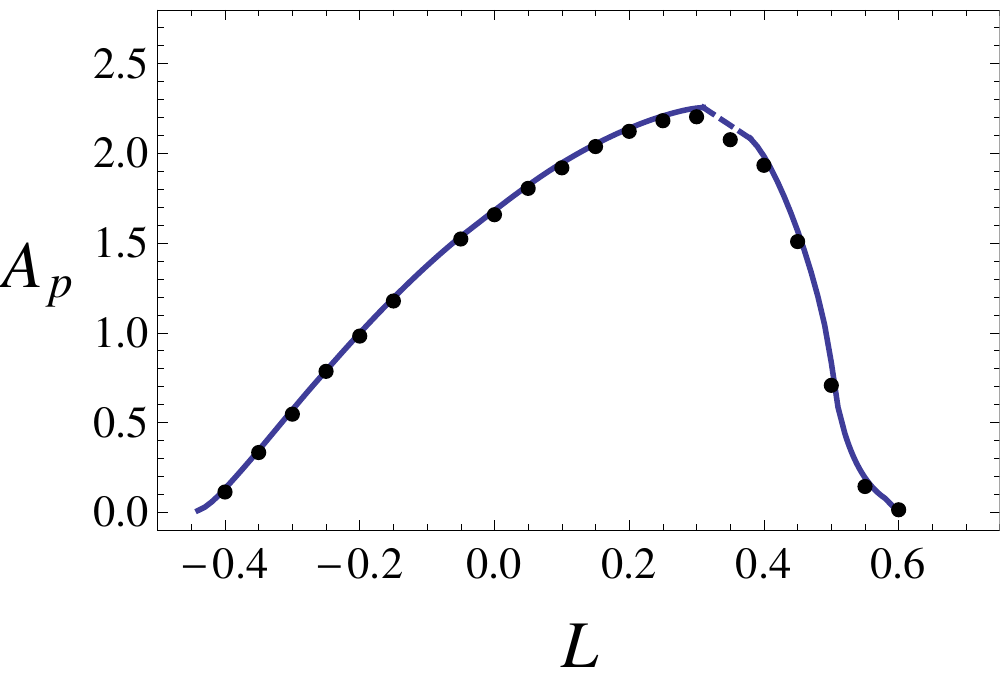}
	\label{fig: area_08}
	\end{subfigure}

	\begin{subfigure}{0.3\textwidth}
	\centering
	\caption{$h=-0.7$}
	\includegraphics[width=\textwidth]{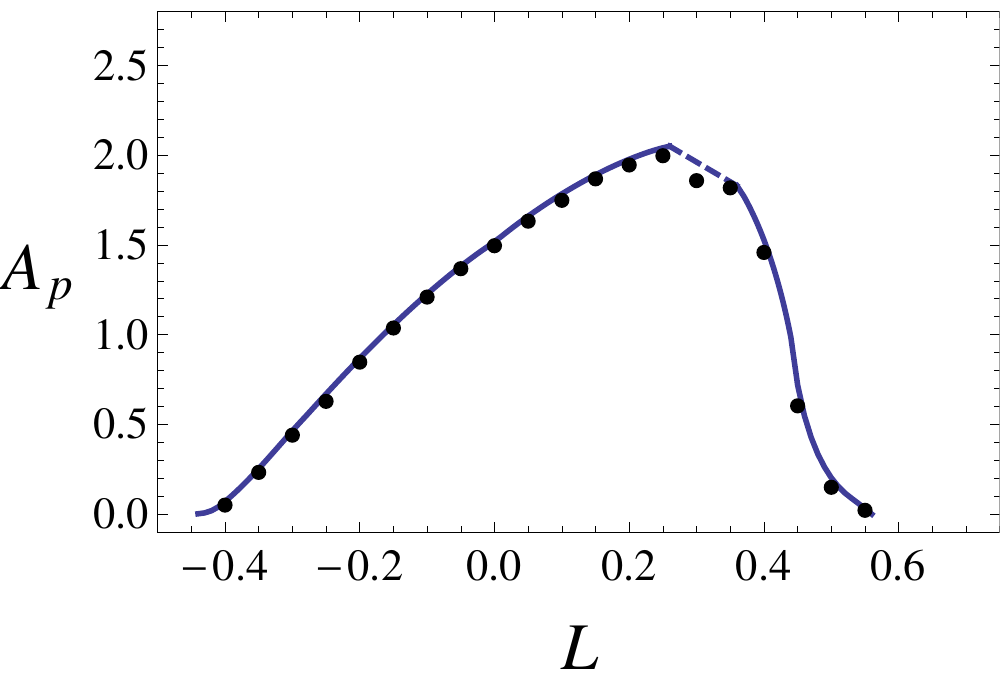}
	\label{fig: area_07475}
	\end{subfigure}	
	\begin{subfigure}{0.3\textwidth}
	\centering
	\caption{$h=-0.6$}
	\includegraphics[width=\textwidth]{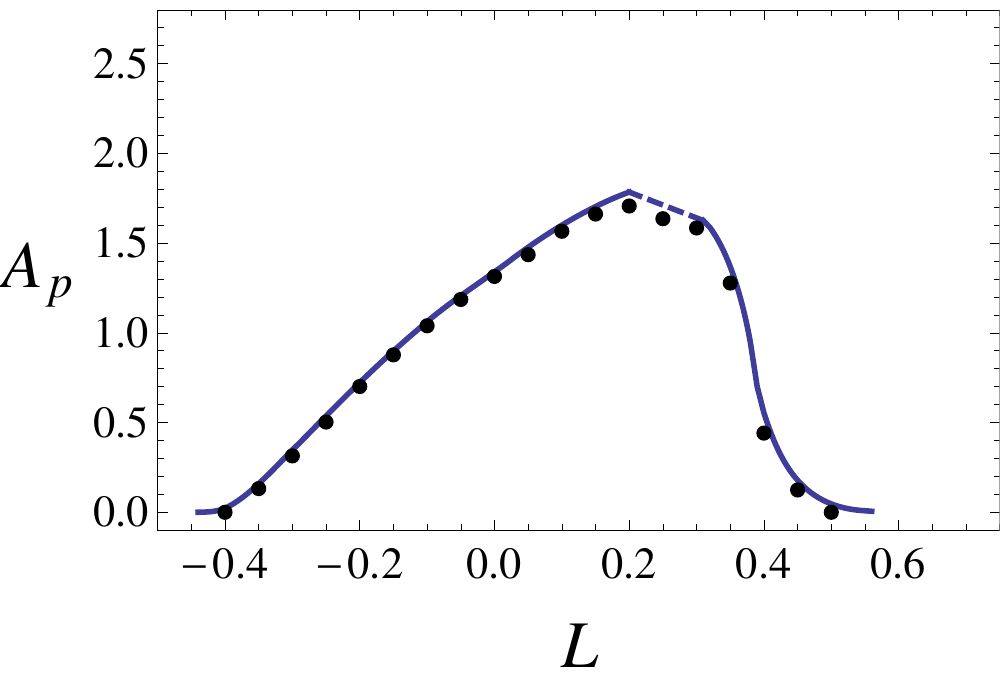}
	\label{fig: area_06}
	\end{subfigure}
	\begin{subfigure}{0.3\textwidth}
	\centering
	\caption{$h=-0.5$}
	\includegraphics[width=\textwidth]{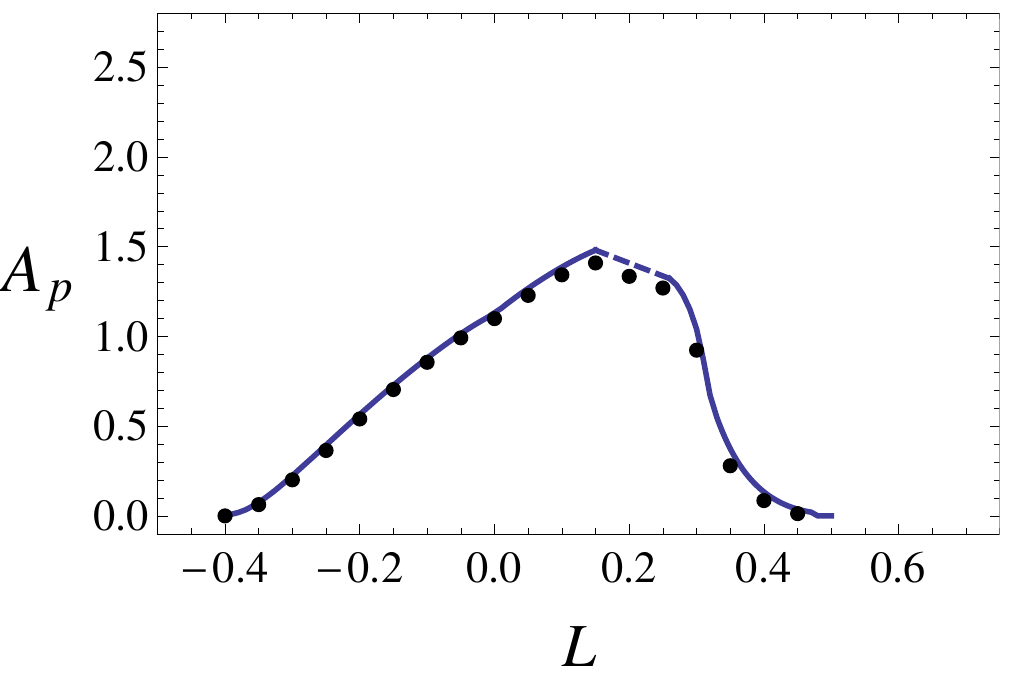}
	\label{fig: area_05}
	\end{subfigure}
\caption{Results similar to the ones of \autoref{fig:area_h_0_7475_nt2} for various values of $h$.}
\label{fig:area}
\end{figure}

The calculations of the permitted areas of motion which are shown in \autoref{fig:area} have been done in a similar way as the one described in \autoref{area}. As we can see, the curves in all panels follow a similar pattern. For $L \lesssim -0.4$ no orbits exist, as no initial conditions satisfy both the $h$ and $L$ restrictions. As $L$ increases (up to $L\lesssim 0.15-0.35$ depending on $h$), the size of the permitted area grows larger as the $J_{1_{up}(R_3=0)}$ curve moves upwards in \autoref{fig: sections_scans}. For large values of $L$ ($L\gtrsim 0.25-0.45$ depending on $h$) the boundaries are defined by the constraint $R_2=0$. As $L$ increases even further, the permitted area shrinks because the boundaries defined by $J_{1_{up}(R_2=0)}$ and $J_{1_{lo}(R_2=0)}$ come closer. The dashed line corresponds to the intermediate values of $L$, where the ambiguity in the calculation of $A_p(L)$ occurs (as we also did in \autoref{fig:area_h_0_7475_nt2}). Even for these values of $L$ this rough approximation is again very close with the results obtained by the scan maps. 

By examining the sequence of plots of \autoref{fig:area}(a)-(f) we conclude that the total area of permitted orbits decreases as $h$ increases.

\paragraph{Chaoticity percentages.}

\begin{figure}[h]
	\centering
	\begin{subfigure}{0.3\textwidth}
	\centering
	\caption{$h=-1.1$}
	\includegraphics[width=\textwidth]{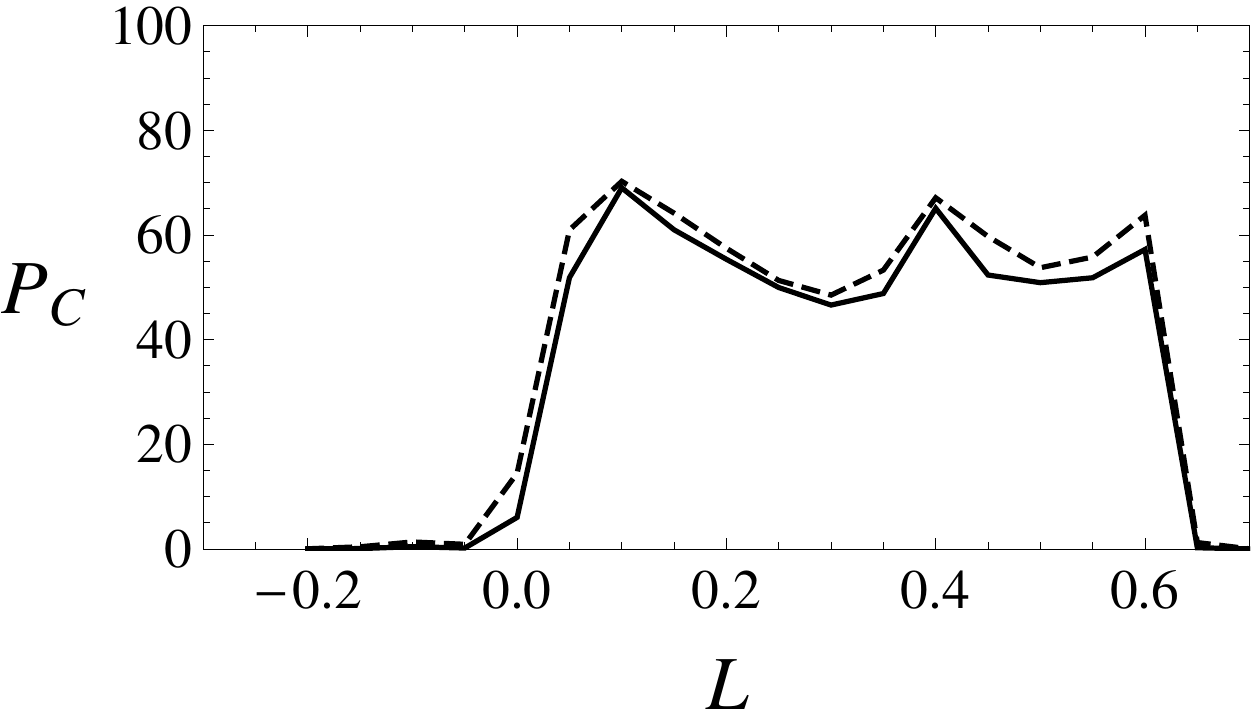}
	\label{fig: pcnt_11}
	\end{subfigure}
	\begin{subfigure}{0.3\textwidth}
	\centering
	\caption{$h=-0.9$}
	\includegraphics[width=\textwidth]{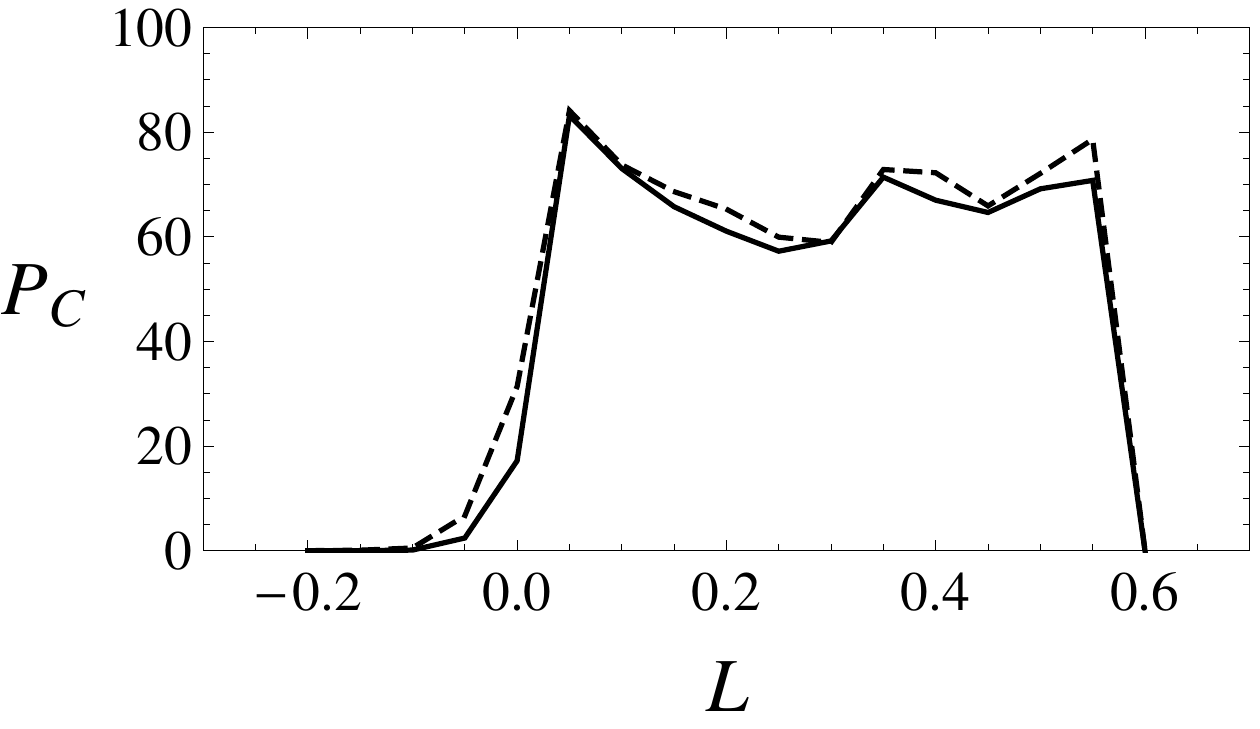}
	\label{fig: pcnt_09}
	\end{subfigure}
	\begin{subfigure}{0.3\textwidth}
	\centering
	\caption{$h=-0.8$}
	\includegraphics[width=\textwidth]{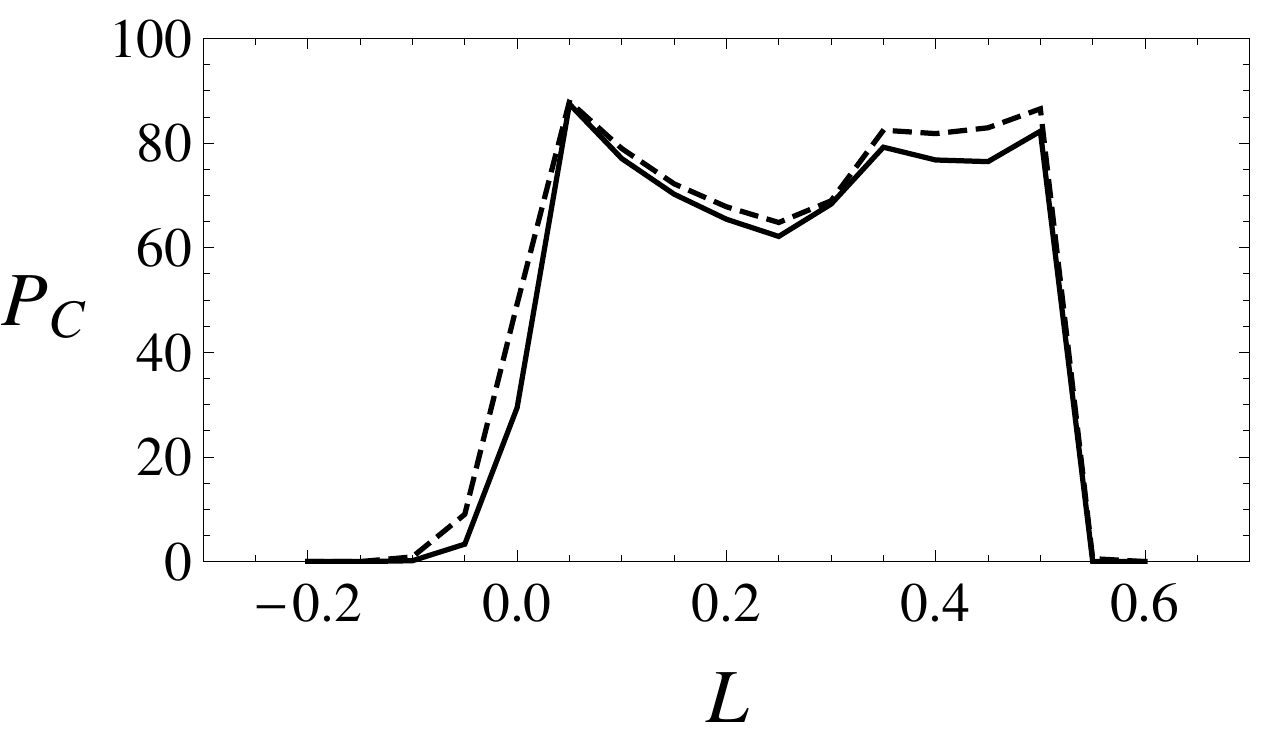}
	\label{fig: pcnt_08}
	\end{subfigure}

	\begin{subfigure}{0.3\textwidth}
	\centering
	\caption{$h=-0.7$}
	\includegraphics[width=\textwidth]{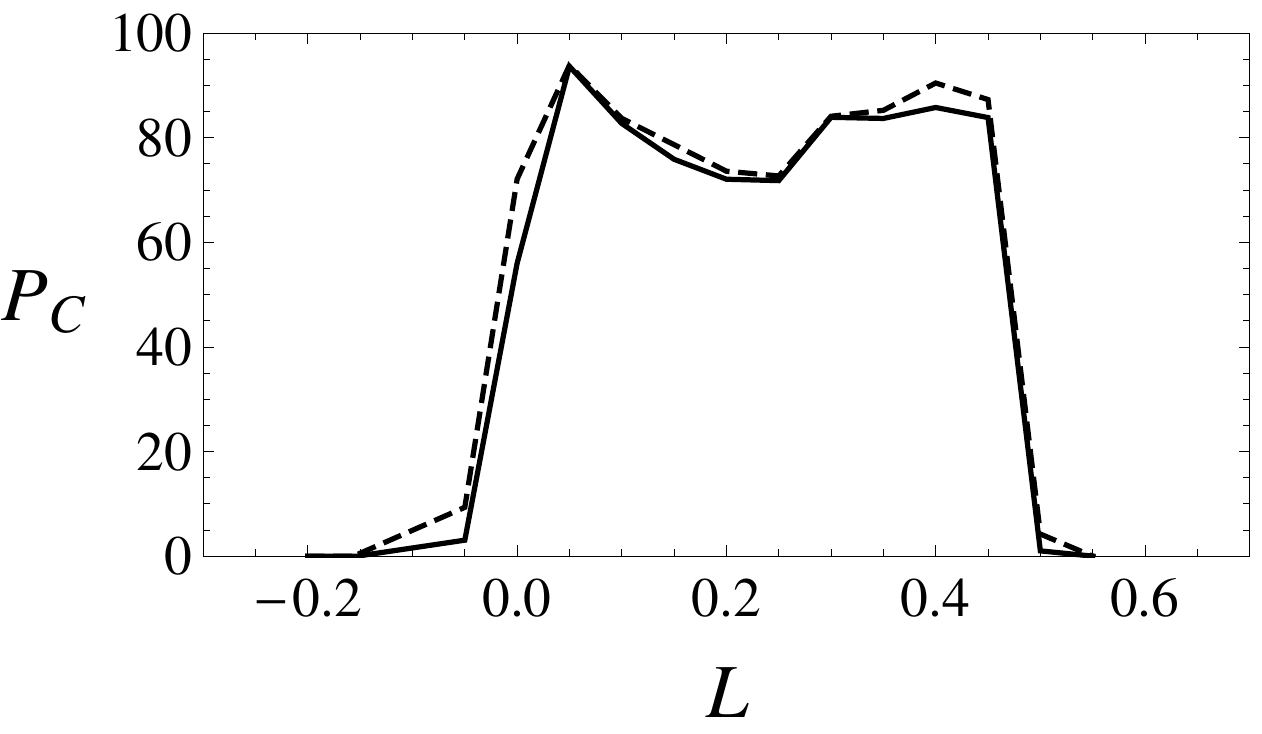}
	\label{fig: pcnt_07}
	\end{subfigure}	
	\begin{subfigure}{0.3\textwidth}
	\centering
	\caption{$h=-0.6$}
	\includegraphics[width=\textwidth]{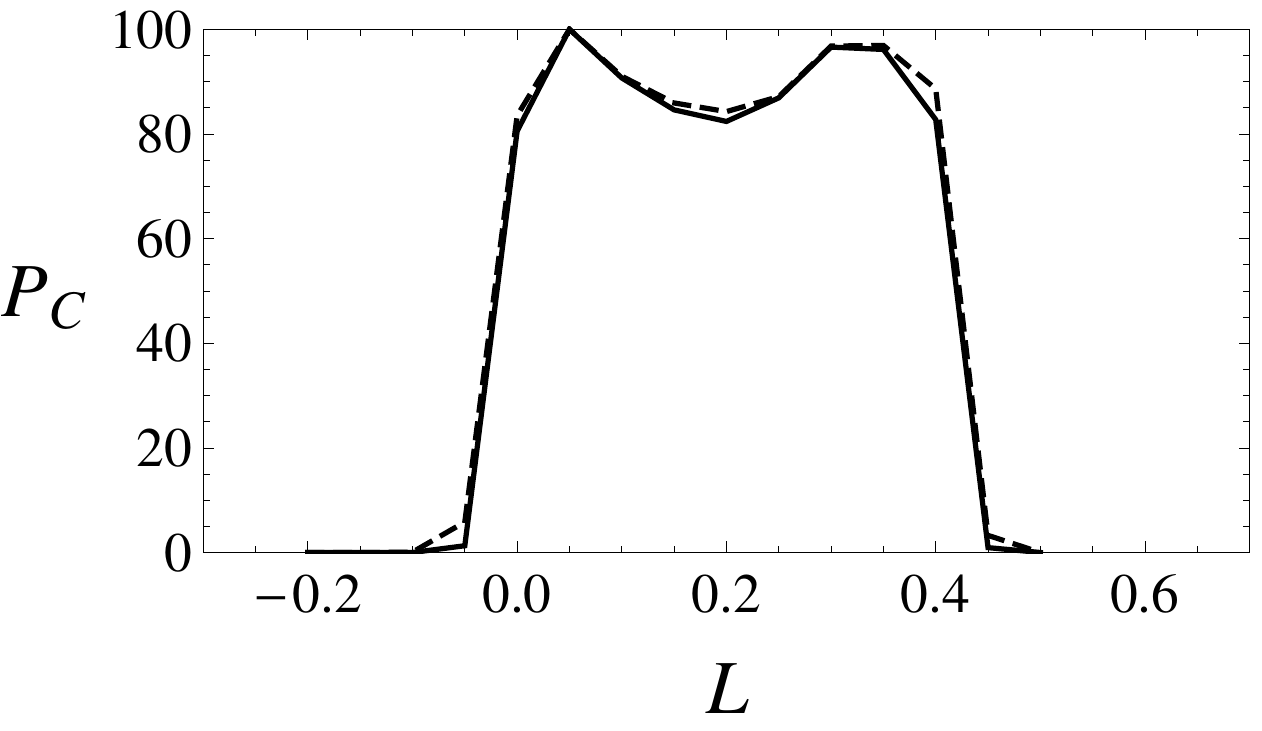}
	\label{fig: pcnt_06}
	\end{subfigure}
	\begin{subfigure}{0.3\textwidth}
	\centering
	\caption{$h=-0.5$}
	\includegraphics[width=\textwidth]{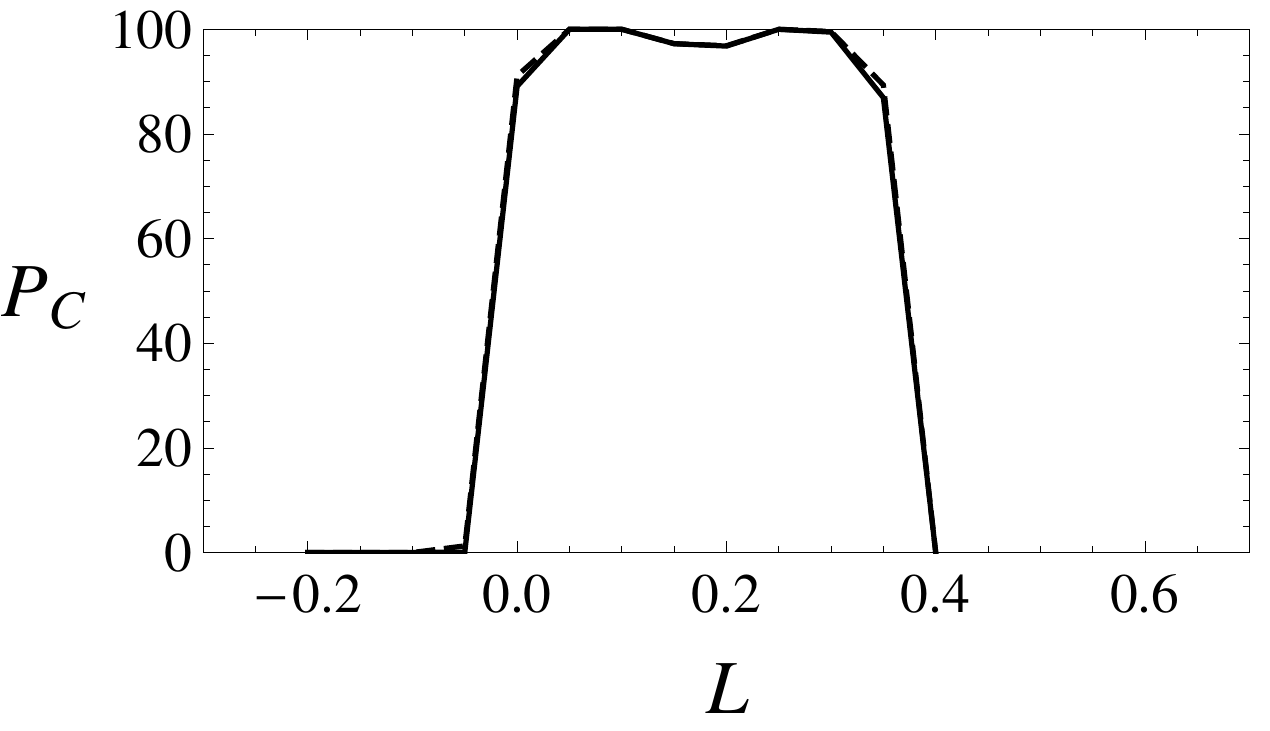}
	\label{fig: pcnt_05}
	\end{subfigure}
\caption{Results similar to the ones of \autoref{fig:pcnt_07475_both} for various values of $h$.}
\label{fig: pcnt}
\end{figure}

We calculate the percentage, $P_C$, of initial conditions leading to chaotic motion, within the set of the permitted initial conditions of the corresponding grid for $-1.1\lesq h\lesq-0.5$ and $-0.2\lesq L\lesq 0.6$, setting $t_{max}=500$. The obtained results are plotted in \autoref{fig: pcnt} by solid lines. Again, resetting the final integration time to $t_{max}=3000$ (dashed curves in \autoref{fig: pcnt}) no significant differences are observed.
The percentages slightly increase because some `sticky' chaotic orbits are now characterized as chaotic, but apart from that the obtained curves are very close to the ones constructed for $t_{max}=500$.

There seems to exist the same trend for all different values of the energy as already discussed in section IV.A.2 for $h=-0.7475$. For small values of the angular momentum ($L\lesssim -0.2$) no chaotic orbits exist. As $L$ increases, $P_C$ increases rather quickly and after an interval where it retains considerably high values,
 it drops down. Finally $P_C$ vanishes for large values of $L$ ($L\gtrsim 0.4-0.6$ depending on $h$), while at the same time the number of permitted initial conditions in the phase space shrinks (\autoref{fig:area}). The values of $L$ for which the chaotic region appears (denoted $L_c$) and disappears (denoted $L_d$) depend on the particular value of the energy $h$. It is interesting to note that within this interval there is at least one local minimum, implying a local maximum in the fraction of regular trajectories.

Our results show that the range $L_d-L_c$ is larger for lower energies. This is related to the size of the permitted PSS area, which is also larger when $h$ is smaller (\autoref{fig:area}). The appearance of the chaotic region seems to happen at about the same value $L\approx{}-0.1$, for all $h$ values, but the eventual shrinking and disappearance of this region varies from $L_d\approx{}0.4$ for $h=-0.5$ (\autoref{fig: pcnt}(f)) to $L_d\approx{}0.7$ for $h=-1.1$ (\autoref{fig: pcnt}(a)).
Presumably, the invariance of the onset of chaoticity is because of the weak $h$
dependence on the critical value $L_c$ for which the central stable periodic orbit undergoes a pitchfork bifurcation. This bifurcation generates the x-shaped chaotic region (\autoref{fig: sections_scans}(b)) which leads to the onset of chaoticity.

In addition, we observe smaller percentages of chaotic motion altogether for lower energies. While the maximum percentage for $h=-0.5$ is $\approx{}100\%$ (\autoref{fig: pcnt}(f)), the one for $h=-1.1$ is just $\approx{}70\%$ (\autoref{fig: pcnt}(a)). This behavior can be explained as follows. As we have seen in \autoref{fig:area}, the higher the energy of the vortices, the smaller the permitted area of motion becomes. Consequently, the orbits of the vortices come closer and the interaction among all three of them becomes stronger, which in turn leads to the enhancement of the chaotic behavior.

We also observe in the panels of \autoref{fig: pcnt} that the `secondary' local maximum between the two `main' local maxima is more pronounced for low energies, for example it has the same height as the two `main' maxima for $h=-1.1$ (\autoref{fig: pcnt}(a)), while it becomes less distinct as the value of $h$ increases, and practically disappears for $h=-0.5$ (\autoref{fig: pcnt}(f)). This happens because, as the value of $h$ increases, the overall percentages of the chaotic orbits increase. Consequently, this phenomenon becomes less significant and eventually not observable.

We believe that this analysis, based on the SALI,
offers a systematic view of the PSS and
the fraction of accessible orbits in it (as per \autoref{fig:area}), as
well as of the fraction of chaotic orbits in it (as per \autoref{fig: pcnt})
and how these change as a function of the canonical physical
properties of the system, namely its energy and its angular
momentum.

\section{Conclusions - Future Directions}

In the present work, we explored a theme of current interest within the
research of atomic BECs, namely the recently realized experimentally
tripoles of vortices and their associated nonlinear dynamical evolution.
We found that this Hamiltonian system is arguably prototypical
(at least within the realm of isotropic magnetic traps) in its exhibiting
chaotic dynamics as parameters or initial conditions are varied. We
focused here on the variation of initial conditions, through the
variation of associated conserved quantities such as the energy
and the angular momentum. Our aim was to associate a technique that
has been previously used in a variety of other low dimensional
settings, namely the SALI diagnostic, for efficiently measuring the chaoticity of the
orbits within this atomic physics realm of vortex dynamics under their
mutual interactions and their individual precession within the
parabolic trap. We found that the SALI is a very accurate diagnostic
of the different levels of chaoticity of the system and enables a qualitative understanding of
how this chaoticity changes as the conserved quantities
are varied, as well as a quantification of the chaotic fraction of
the phase space of the system.

This work paves the way for the consideration of a wide range
of additional problems within the dynamics of coherent structures
in the realm of Bose-Einstein condensates.
First of all, it would be straightforward to explore how
the dynamics of this tripole would compare/contrast to the
recently explored~\cite{dsh3} dynamics of 3 co-rotating
vortices (i.e., vortices of the same charge). Another natural
extension in the vortex case would be to examine how the
chaotic region expands as a fourth vortex of either a positive
or a negative charge comes into play. The special cases of
4 co-rotating vortices (with relevant square and rhombic etc.
stable configurations), as well as the case of the generally
fairly robust~\cite{middel10} vortex quadrupole would be
of interest in this setting. Additionally, extending such
considerations to other dimensions would present interesting
possibilities as well. On the one hand,
a wide range of theoretical and experimental considerations (including
particle based approaches, such as the ones
utilized herein) have been developed for dark
solitons in 1d; see e.g. the recent review of Frantzeskakis~\cite{djf}.
On the other hand, generalizing to 3 dimensions and the consideration
of multiple vortex rings and their dynamics~\cite{komineas}
would be equally or even more exciting from the point of
view
of ordered vs. chaotic dynamics. Examination of these
directions is currently in progress and will be reported
in future publications.

\acknowledgments
\noindent The authors would like to thank Prof. Roy H. Goodman for a comment that provided useful insights toward further development of this direction.

\noindent N.K.~acknowledges support from the MC Career Integration Grant PCIG13-GA-2013-618399.

\noindent V.K.~and Ch.S.~have been co-financed by the European Union (European Social Fund - ESF) and Greek national funds through the Operational Program ``Education and Lifelong Learning'' of the National Strategic Reference Framework (NSRF) - Research Funding Program: THALES. Investing in knowledge society through the European Social Fund. Ch.S.~was also supported by the Research Committees of the University of Cape Town (Start-Up Grant, Fund No 459221) and the Aristotle University of Thessaloniki (Prog.~No 89317).

\noindent P.G.K.~acknowledges support from the National Science Foundation
under grants CMMI-1000337, DMS-1312856, from the Binational Science
Foundation under grant 2010239, from FP7-People under grant IRSES-606096
and from the US-AFOSR under grant
FA9550-12-10332.

%

\end{document}